\documentclass[useAMS,usenatbib]{mn2e}
\usepackage{amsmath}
\usepackage{amssymb}
\usepackage{graphicx}

\usepackage{scrextend}

\usepackage{url}

\newcommand*\aap{A\&A}

\newcommand*\aj{AJ}

\newcommand*\apj{ApJ}
\newcommand*\apjl{ApJ}

\newcommand*\apjs{ApJS}

\newcommand*\araa{ARA\&A}

\newcommand*\mnras{MNRAS}

\usepackage[]{hyperref}   
\hypersetup{bookmarks=true,colorlinks=true,linkcolor=black,citecolor=black,urlcolor=blue}

\newcommand{\dw}{2M1821$+$14}

\title[High-precision parallax of \dw\ with GTC]{Parallax of the L4.5 dwarf \dw\ from high-precision astrometry with OSIRIS at GTC\thanks{Based on observations made with the Gran Telescopio Canarias (GTC), installed at the Spanish Observatorio del Roque de los Muchachos of the Instituto de Astrof\'isica de Canarias, in the island of La Palma (Program IDs GTC37-13A and GTC11-14A).}}
\author[Sahlmann et al.]
{J. Sahlmann$^{1}$\thanks{E-mail: Johannes.Sahlmann@esa.int}\thanks{ESA Research Fellow}, 
P. F. Lazorenko$^{2}$, 
H. Bouy$^{3}$,
E. L. Mart\'in$^{4}$,
D. Queloz$^{5,6}$,\newauthor
D. S\'egransan$^{6}$,
M. R. Zapatero Osorio$^{4}$
\\
$^{1}$European Space Agency, European Space Astronomy Centre, P.O. Box 78, Villanueva de la Ca\~nada, 28691 Madrid, Spain\\
$^{2}$Main Astronomical Observatory, National Academy of Sciences of the Ukraine, Zabolotnogo 27, 03680 Kyiv, Ukraine\\
$^{3}$INTA-CSIC Centro de Astrobiolog\'ia, Departamento de Astrof\'isica, P.O. Box 78, Villanueva de la Ca\~nada, 28691 Madrid, Spain\\ 
$^{4}$INTA-CSIC Centro de Astrobiolog\'ia, 28850 Torrej\'on de Ardoz, Madrid, Spain\\     
$^{5}$University of Cambridge, Cavendish Laboratory, J J Thomson Avenue, Cambridge, CB3 0HE, UK  \\
$^{6}$Observatoire de Gen\`eve, Universit\'e de Gen\`eve, 51 Chemin Des Maillettes, 1290 Versoix, Switzerland}

\date{Accepted 2015 September 28. Received 2015 September 28; in original form 2015 March 13}

\pagerange{\pageref{firstpage}--\pageref{lastpage}} \pubyear{2015}
\begin{document}

\label{firstpage}
\maketitle

\begin{abstract}
We used the OSIRIS camera at the 10.4 m Gran Telescopio Canarias (GTC) to monitor the astrometric motion of the L4.5 dwarf \dw\ over 17 months. The astrometric residuals of eleven epochs have a r.m.s. dispersion of 0.4 mas, which is larger than the average precision of 0.23 mas per epoch and hints towards an additional signal or excess noise. Comparison of the point-spread-functions in OSIRIS and FORS2/VLT images reveals no differences critical for high-precision astrometry, despite the GTC's segmented primary mirror. We attribute the excess noise to an unknown effect that may be uncovered with additional data.

For \dw, we measured a relative parallax of $106.15 \pm 0.18$ mas and determined a correction of $0.50\pm0.05$ mas to absolute parallax, leading to a distance of $9.38 \pm0.03$ pc. We excluded at 3-$\sigma$ confidence the presence of a companion to \dw\ down to a mass ratio of 0.1 ($\approx 5\, M_\mathrm{Jupiter}$) with a period of 50--1000 days and a separation of 0.1--0.7 au. The accurate parallax allowed us to estimate the age and mass of \dw\ of 120--700 Myr and 0.049$^{+0.014}_{-0.024}$ M$_\odot$, thus confirming its intermediate age and substellar mass. We complement our study with a parallax and proper motion catalogue of 587 stars ($i'\simeq15.5-22$) close to \dw, used as astrometric references. This study demonstrates sub-mas astrometry with the GTC, a capability applicable for a variety of science cases including the search for extrasolar planets and relevant for future astrometric observations with E-ELT and TMT.
\end{abstract} 

\begin{keywords}
astrometry --  brown dwarfs -- technique: high angular resolution -- atmospheric effects  -- parallaxes 
\end{keywords}

\section{Introduction}
High-precision astrometry better than one milli-arcsecond (mas) leads to key scientific results in the fields of, e.g., galaxy kinematics \citep{Sohn}, galactic centre dynamics \citep{Gillessen}, binary stars \citep{Tokovinin}, brown dwarfs \citep{Lane:2001le}, and extrasolar planets \citep{Benedict, Sahlmann:2011}, but only few instruments provide us with such exquisite accuracy. Using the FORS2 camera at the ESO/VLT, we demonstrated that large-aperture ground-based telescopes offer the possibility to obtain 100 micro-arcsecond  astrometry over several years with a field of view of 2--4\arcmin\ \citep{Lazorenko2009,  Lazorenko:2011lr}. These capabilities are already used for a planet search around ultra-cool dwarfs since 2010 \citep{Sahlmann:2014aa},  sensitive to companions as light as Neptune in 500-2000 day orbits, a parameter space difficult to  explore by radial velocity or transit photometry searches. Accompanying results include the measurement of ultracool dwarf parallaxes at the unprecedented level of 0.1\%. 

Here, we explore the astrometric capabilities of OSIRIS imaging observations. Because OSIRIS/GTC is similar to FORS2/VLT in terms of telescope aperture size, field of view, pixel scale, and bandpass, we expect to also achieve comparable astrometric performances. However, the most prominent difference between VLT and GTC is the latter's segmented primary mirror. It is not known how the segmented mirror and the associated instrumental point-spread-function (PSF) affect the achievable astrometric precision, which for FORS2 corresponds to photocentre measurements at the milli-pixel level. Our study explores a field that is relevant for the Extremely Large Telescopes (ELT), all based on segmented primary mirrors, which theoretically enable accuracies at the 10 micro-arcsecond ($\mu$as) level \citep{Lazorenko2009, Trippe}.

As a test case for OSIRIS, we monitored the astrometric motion of one nearby brown dwarf over several epochs. The goal of this paper is threefold: 1) determine the astrometric performance of OSIRIS/GTC; 2) measure the parallax of \dw\ and search for orbiting companions; 3) estimate the potential of OSIRIS for an astrometric planet search targeting optically faint objects. 

\section{Observations}
\begin{table*}
\caption{Observation record for OSIRIS/GTC imaging of \dw\ with the Sloan $i'$ filter.}
\centering
\begin{tabular}{rrrrrrcc}
\hline
No. & Mean date & $N_\mathrm{f}$ & $\Delta t$ & Air- & FWHM & Used & Comment / Program\\
 & (UT) &  & (h) & mass& (\arcsec) \\
\hline
1 &  2013-05-08T03:55:25 & 29 & 0.88  &1.04&  0.93 &  No& Insufficient image quality\\
2 & 2013-05-20T01:30:47& 26 & 0.92 & 1.20 & 0.79 & Yes &GTC37-13A\\
3 & 2013-06-03T04:32:46& 25 & 0.84 & 1.14 & 0.65 & Yes &GTC37-13A\\
4 & 2013-06-05T03:25:29& 24 & 0.84 & 1.05 & 0.92 & Yes &GTC37-13A\\
5 & 2013-07-04T04:39:58& 25 & 0.77 & 1.81 & 0.72 & Yes &GTC37-13A\\
6 & 2013-07-05T02:45:32& 27 & 0.84 & 1.19 & 0.63 & Yes &GTC37-13A\\
7 & 2013-07-12T00:35:07&  27 & 1.02  &1.04 & 0.72  & No&Pointing off by $\sim$11\arcsec\\
8 & 2013-07-31T23:03:41& 55 & 1.82 & 1.04 & 0.79 & Yes &GTC37-13A\\\
9 & 2013-08-16T00:34:53 & 28 & 0.87  &1.30 & 0.75  & No&Pointing off by $\sim$11\arcsec\\
10 & 2014-07-05T23:59:15& 24 & 1.07 & 1.05 & 0.83 & Yes &GTC11-14A\\
11 & 2014-08-05T00:18:31& 25 & 0.74 & 1.12 & 0.65 & Yes &GTC11-14A\\
12 & 2014-08-31T22:10:17& 20 & 0.57 & 1.09 & 0.92 & Yes &GTC11-14A\\
13 & 2014-09-11T22:12:43& 24 & 0.74 & 1.18 & 0.84 & Yes &GTC11-14A\\
14 & 2014-10-07T21:18:03& 26 & 0.74 & 1.35 & 0.64 & Yes &GTC11-14A\\
15 &  2013-04-09T04:42:26 & 35 & 0.92 & 1.13&  0.65   & No& Pointing off by $\sim$2\arcmin\\
16 & 2013-05-03T03:18:21 & 29 & 0.87  &1.11 & 0.93   & No& Pointing off by $\sim$2\arcmin\\
17 & 2014-06-03T00:24:54 & 29&  0.80  &1.04 & 0.92   & No& Pointing off by $\sim$2\arcmin\\
\hline
\end{tabular}
\label{tab:obs}
\end{table*}
We searched the target for this study in the \url{DwarfArchives.org} list of M, L, and T dwarfs, where we considered only sources with previously unknown parallax and that fulfil the observational requirements set essentially by the target's optical magnitude and reference star density and the telescope location. We selected the L dwarf \href{http://simbad.u-strasbg.fr/simbad/sim-basic?Ident=2MASS+J18212815\%2B1414010&submit=SIMBAD+search}{{2MASS J18212815+1414010}}, hereafter \dw, which was discovered by \cite{Looper:2008aa} in a 2MASS proper motion survey. At a galactic latitude of $\sim$$13\degr$, \dw\ is located close to the plane and therefore in a region replete with background stars. \cite{Looper:2008aa} classified it as L4.5 in the optical, noting spectral features that indicated an unusually dusty atmosphere and/or youth. They also discovered Li I absorption and estimated a spectrophotometric distance of $\sim$10 pc.  \cite{Kirkpatrick:2010lr} present radial velocity and proper motion measurements and \citep{Blake} presented two-epoch ($\Delta T = 162$ d) radial velocity measurements, showing no sign of close binarity. Despite being included in several other studies \citep[e.g.]{Yang:2015aa,Metchev:2015aa}, no trigonometric parallax measurement of \dw\ was yet published, which probably is due to its relatively faint optical \citep[$I_\mathrm{C}=17.0$,][]{Koen:2013uq} and infrared \citep[$J=14.43$,][]{Skrutskie:2006fk} magnitude.

\begin{figure} 
\includegraphics[width=\linewidth]{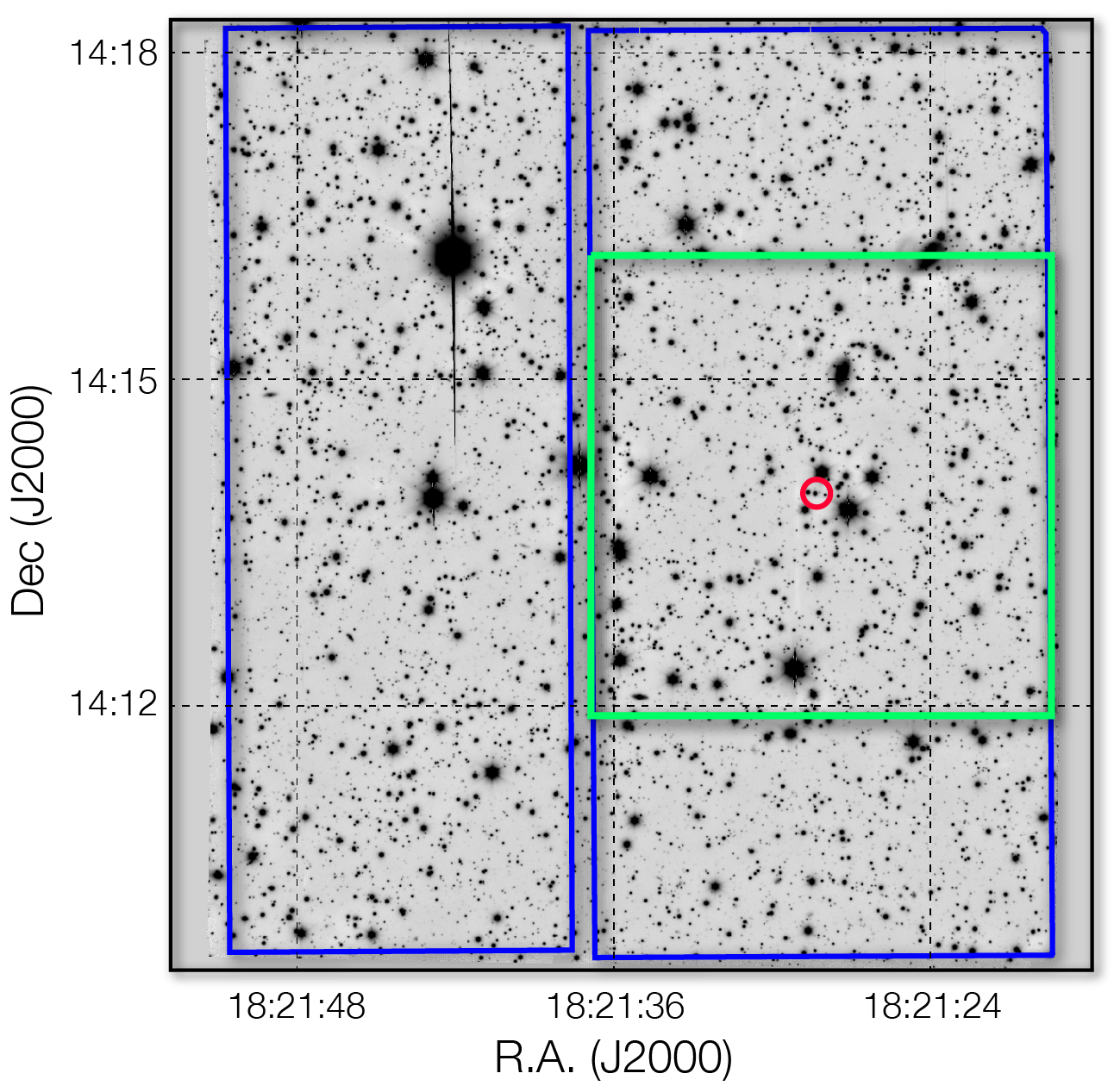}
\caption{Field of view of OSIRIS. The background image was obtained by stacking all exposures of our programme. The unvignetted footprints of CCD1 (left) and CCD2 (right) are indicated in blue. The $4\arcmin\!\times\!4\arcmin$ area of CCD2 used for astrometry is delineated in green and centred on the target \dw\ (encircled).}
\label{fov}
\end{figure}

We observed \dw\ with the OSIRIS camera \citep{Cepa:2003kx} of the GTC in the broad-band $i'$-filter with an exposure time of 45 s and $1\times1$~pixel binning. The camera images a $\sim$\,$8\arcmin\!\times\!8\arcmin$ field of view on two detector chips, CCD1 and CCD2. To avoid effects related to camera instabilities and relative chip motion, which we observe on FORS2 \citep{Lazorenko:2014aa}, we used only CCD2 for the astrometry and position the target at its centre, see Fig. \ref{fov}. Over a timespan of 506 d ($\sim$1.4 y) from May 2013 to October 2014 we obtained 17 epochs, each consisting of about $N_\mathrm{f}=25-30$ individual exposures spanning $\Delta t=55$ min on average. A $\pm 1$\arcsec\ jitter was applied between consecutive exposures. 

Table \ref{tab:obs} summarises the observations, including the epoch No., the mean date of the epoch exposures, the average airmass, the average FWHM measured for star images, and indicates whether the epoch was used to produce the final results. One epoch has insufficient image quality and for three epochs the target was positioned at the CCD2 edge. In two more epochs, the target was not centred sufficiently well in CCD2 and these images were not used either in the final reduction (see Sect.\,\ref{ep_res}). The results were thus obtained with the data of eleven epochs. 

\section{Data  reduction and image analysis}{\label{im_red}}
The raw data were bias-subtracted and flatfielded using standard methods and the calibration data provided to us by the observatory. For the astrometric analysis we used the 2000$\times$2000 pixel central area of CCD2, which with the nominal 0\farcs127 pixel scale measures $\sim\!4\arcmin\!\times \!4$\arcmin. In this area we detected more than 800 unsaturated stars brighter than $i'=22$ and measured the positions of their photocentres. Considering that the global instrumental parameters (telescope aperture, pixel scale, field of view) of OSIRIS/GTC are similar to FORS2/VLT, we  performed the astrometric reduction with the same methods that ensure the 0.1~mas astrometric precision for FORS2 data. This high astrometric precision is reached due to effective averaging of the atmospheric image motion over the large telescope aperture. The details and the latest refinements of the method are described in \citet{Lazorenko:2014aa}.

\subsection{Comparison of the OSIRIS and FORS2 image profiles}{\label{comp}}
One goal of this study is the investigation of the impact of the segmented mirror structure on the astrometric performance. Therefore we analysed the shapes of star images obtained with OSIRIS and FORS2 in the central area (Sect. \ref{profiles}) and in the wings (Sect. \ref{wings}). In Sect.\,\ref{fwhm_size} we then compare the FWHM and the intensity of the atmospheric image motion registered on both instruments.

\subsubsection{PSF kernel profile}{\label{profiles}}
The star photocentres $\bar{x}$, $\bar{y}$ were computed in the same way as for FORS2 \citep{Lazorenko2006} by fitting the measured  counts $p_{i,j}$ 
for the pixel $i$, $j$ with the two-dimensional model of the PSF
\begin{equation}
\label{eq:psf}
P(x,y)=G(x,y)+(x-\bar{x})^2 G'(x,y) +(y-\bar{y})^2 G''(x,y), 
\end{equation}
which includes the principal elliptical Gaussian $G(x,y)$, {where the inclination relative to the coordinate axes is a free parameter}, and two auxiliary Gaussians $G'(x,y)$ and $G''(x,y)$, centred at $\bar{x}$, $\bar{y}$ and with semi-axes aligned with the CCD axes. The model contains 12 free parameters {that vary depending on the star's position in the field of view (FoV), the seeing, and the telescope optics adjustment}. Conversion from the continuous function $P(x,y)$ to the discrete counts $\hat P_{i,j}$ for a pixel $i$, $j$, which should be obtained by integration of the function Eq. (\ref{eq:psf}) within a pixel, was implemented in an analytic way. {The model Eq. (\ref{eq:psf}) provides good fits to the data to determine the instant image parameters for a symmetric star profile with a solution that is robust to} small fluctuations of the pixel counts because of the weak correlation between the model parameters and $\bar{x}$, $\bar{y}$ \citep[cf.][]{Lazorenko:2014aa}.

\begin{figure}
\includegraphics[width=\linewidth]{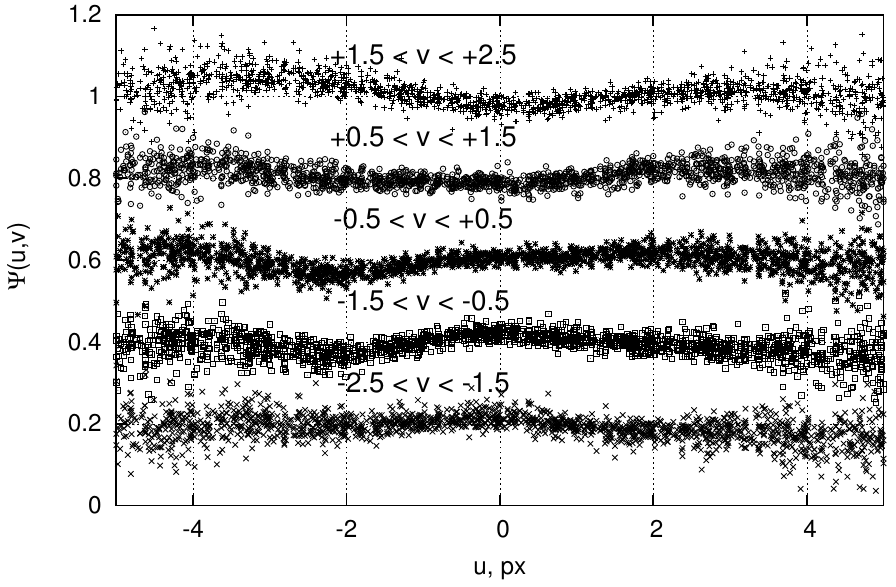}
\caption{Distribution of the normalised residuals $\psi(u,v)$ of the `measured -- model` pixel counts near the PSF kernel  for  a FORS2 example image with FWHM\,$=0\farcs65$.  The horizontal $u$-axis is aligned with the $x$-axis of the CCD and each sequence of data points corresponds to a slice along the $v$-axis. }
\label{psf_fors}
\end{figure}

The  residuals $p_{i,j}-\hat P_{i,j}$ within the fit area  of $11\times 11$~pixel size around the star centres are almost randomly scattered but contain a weak signal which modulates the shape of the PSF. It is hidden in the noise and revealed only if we combine the data of many stars.  In order to characterise it, we computed the normalised residuals 
\begin{equation}
\label{eq:psi}
\psi(u,v) =(p_{i,j}-\hat P_{i,j})/ \hat P_{i,j}
\end{equation}
where  $u=x_i-\bar{x}$ and $v=y_j-\bar{y}$ is the  relative distance between the  centre $x_i$, $y_j$ of the pixel $i$, $j$ and the star image photocentre  $\bar{x}$, $\bar{y}$. 
Figure \ref{psf_fors} shows the distribution of $\psi(u,v)$ computed for sufficiently bright stars and accumulated over the 2000$\times$1000 px FoV of chip1 for an example FORS2 image with visually circular star profiles. We  present this two-dimensional distribution in the form of slices along the $u$-axis, each of which is one pixel wide on $v$. The centres of the slices are set to  $v=-2 \ldots 2$, near the PSF kernel. One can see that the distribution of $\psi(u,v)$ contains systematic wave-like features whose shape is changing between the slices. 

\begin{figure}
\includegraphics[width=\linewidth]{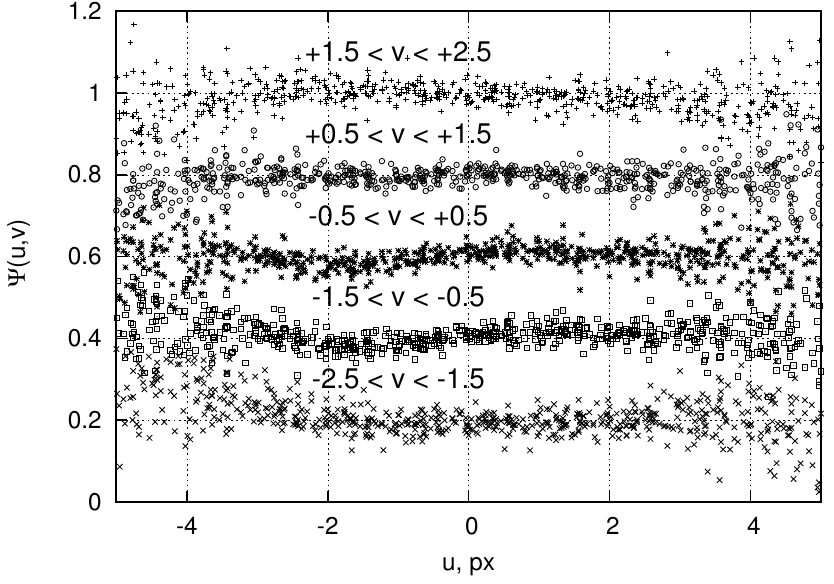}
\includegraphics[width=\linewidth]{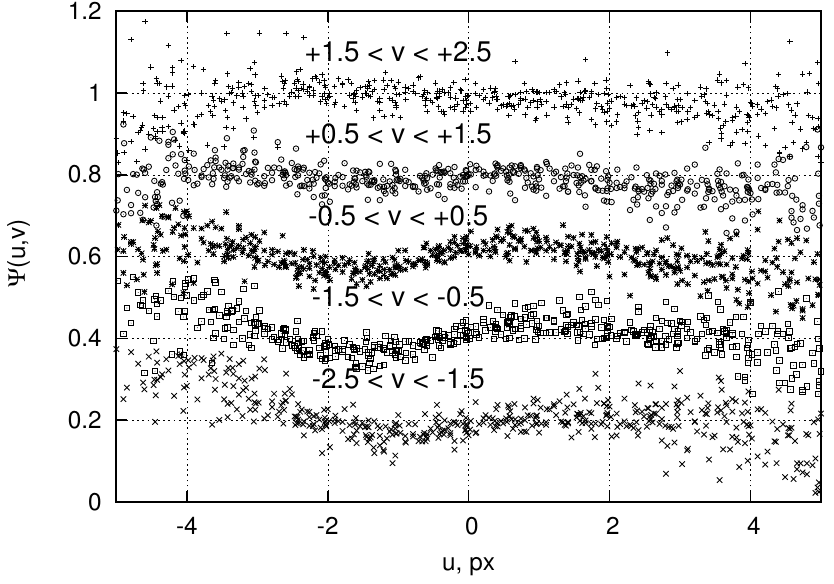}
\caption{The same as Fig.\,\ref{psf_fors} but for an example OSIRIS image. The two panels correspond to the two halves of CCD2 used for astrometric reduction, i.e.\ $x<1000$~px ({\it upper panel}), and  $x>1000$~px ({\it lower panel}).}                    
\label{psf}
\end{figure}

In Fig. \ref{psf} we present the distribution of the residuals $\psi(u,v)$ for OSIRIS separately for the left half, with $x <1000$ pixel (px), and the right  half ($x >1000$~px) of the used area in CCD2. The size of each region is 1000$\times$2000 px, comparable to the FORS2 field in the above analysis, and we chose matching seeing conditions as well. Comparison of Fig.\,\ref{psf_fors} and Fig.\,\ref{psf} shows qualitatively similar wave-like patterns of the function $\psi$ for both telescopes.

\begin{figure}
\includegraphics[width=\linewidth]{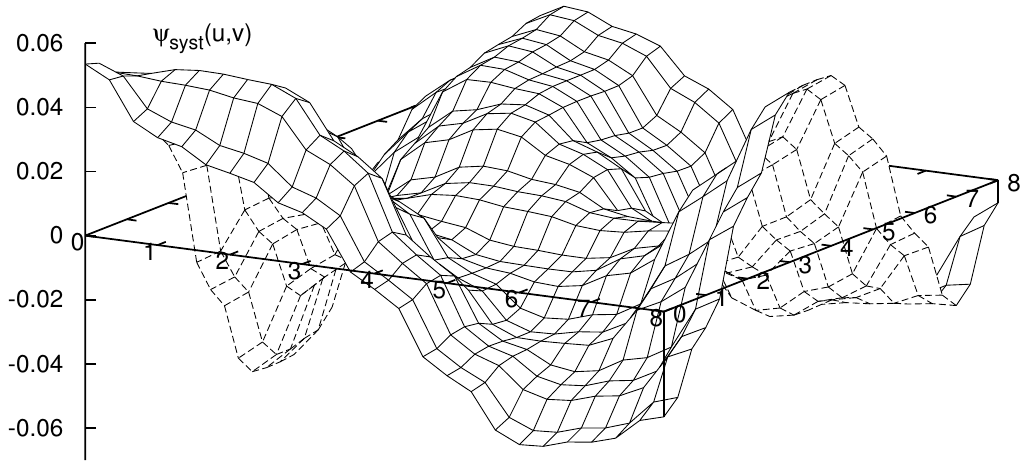}
\caption {Shape of the systematic residuals $ \psi_{\rm syst}(u,v)$   for a series of OSIRIS exposures obtained on 5 July 2013 with FWHM=$0\farcs64$ and visually impeccable images. The r.m.s. of $ \psi_{\rm syst}(u,v)$  within $2\times2$~pixels from the PSF centre is $\sigma(\psi)=0.018$.}
\label{2d}
\end{figure}

The random scatter of $\psi(u,v)$ is caused mainly by photon statistics and therefore increases with $|u|$ because of the decrease of the photoelectron counts at the periphery of the PSF. The systematic part $\psi_{\rm syst}$ of $\psi$  contains two components. The first varies  between the adjacent exposures, thus indicating contribution of the atmospheric turbulence uncorrelated in time. The second  component is stable during the whole frame series of one epoch and is potentially an imprint of the current optical aberrations of the telescope.  An example of the fine structure of $\psi_{\rm syst}$ for OSIRIS measured on 5 July 2013 is illustrated in Fig. \ref{2d}. 

\begin{figure}
\includegraphics[width=\linewidth]{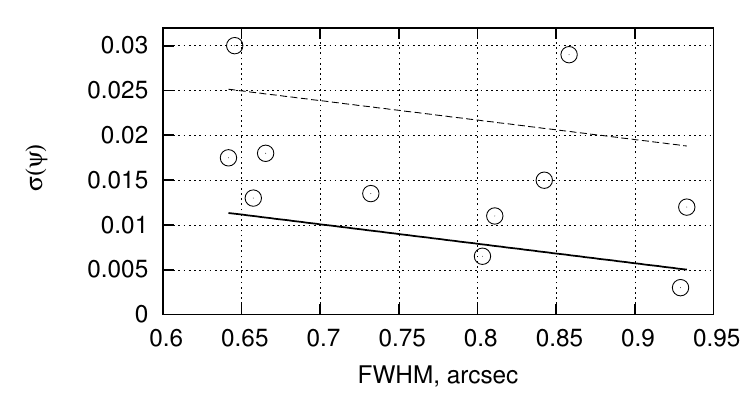}
\caption {The values of $\sigma(\psi)$ as a function of the FWHM  in  OSIRIS images for every epoch (open circles). The equivalent values for FORS2 are below the dashed line in 95\%  cases and their mean value is indicated by the solid line.}
\label{os_f}
\end{figure}

In Fig. \ref{os_f} we present the values of $\sigma(\psi)$, which is the r.m.s. of $\psi_{\rm syst}$ within $\pm 2$~px at the kernel of the PSF.  We compare that to FORS2, for which $\sigma(\psi)$ was obtained for the large dataset of the \cite{Sahlmann:2014aa} survey and is represented here by a the result of a linear fit in FWHM.   In comparison to FORS2, the fluctuations of $\psi_{\rm syst}$ at the PSF centre are on average 50\% larger for OSIRIS. In both cases, the dependence of $\sigma(\psi)$ on FWHM is weak and shows a comparable decreasing trend.

A better representation of the measured counts in the  PSF kernel is given by the model
\begin{equation}
\label{eq:psf2}
PSF(x,y)  =P(x,y)  [1+\psi_{\rm syst}(u,v)], \quad u=x-\bar{x},  \quad   v=y-\bar{y},
\end{equation}
where the analytic function $\psi_{\rm syst}(u,v)$ for each $u$, $v$ is found by smoothing the measured discrete values $\psi(u,v)$ with a  numeric model \citep{Lazorenko2006}. Equation (\ref{eq:psf2}) is solved iteratively by fitting the residuals $p_{i,j} - \hat P_{i,j} \psi_{\rm syst} (u,v)$ with the function $P(x,y)$. We start from the zero approximation $\psi_{\rm syst}=0$. At every iteration  step, we recompute the parameters of $P(x,y)$ and update the function $\psi_{\rm syst}$. We found that the difference $\Delta$ between the initial photocentre positions computed with  $\psi_{\rm syst}=0$, and the final solution for $\bar{x}$, $\bar{y}$ is not constant for different stars. 
For OSIRIS images with $\sigma(\psi) \sim 0.03$,  $\Delta$ varies in the range of 10 to 50 milli-pixel (1--6 mas) and is about 30 milli-pixel on average. The variation of $\Delta$ is caused by the discretised structure of the CCD and depends on the position of the star profile relative to the pixel grid, because that determines the pixel sampling of $\psi_{\rm syst}$. We conclude that the impact of $\psi_{\rm syst}$ cannot be neglected unless it is so small that $\sigma(\psi)$ is below 0.001. In this case the bias in photocentre positions is smaller than 0.1 mas.

The function $\psi_{\rm syst}$ describes the  fine structure of the measured  star profiles  (Eq. \ref{eq:psf2}), which in general are non-symmetric and can have  non-zero  first derivatives of $PSF(x,y)$ on ${x}$ or ${y}$ over significantly large pixel areas. In addition, the shape of the function $\psi_{\rm syst}$ depends on the position in the FoV. However, we were not able to improve precision by modeling these variations because of the limited number of reference stars, thus $\psi_{\rm syst}$ was assumed constant over the FoV.

The complicated structure of the PSF, in particular the one caused by optical aberrations, makes the definition of the `actual` photocentre ambiguous, because it does neither coincide with the model profile centre nor with the weighted photocentre.  The solution $\bar{x}$, $\bar{y}$ obtained using Eq. (\ref{eq:psf2}) can therefore be displaced from the `actual` photocentre. It is difficult to estimate the resulting bias of $\bar{x}$, $\bar{y}$ values, but a priori we can assume that it is proportional to $\sigma(\psi)$, which is the measure of irregularity of $\psi_{\rm syst}$. 
Figure \ref{os_f} shows that with the exception of two cases, the values of $\sigma(\psi)$ for OSIRIS are slightly larger but comparable to those of FORS2. In this respect, the compound structure of the GTC main mirror does hardly affect the astrometry, at least for good images with $\sigma(\psi) < 0.02$.

\subsubsection{PSF wings}
\begin{figure}
\includegraphics[width=\linewidth]{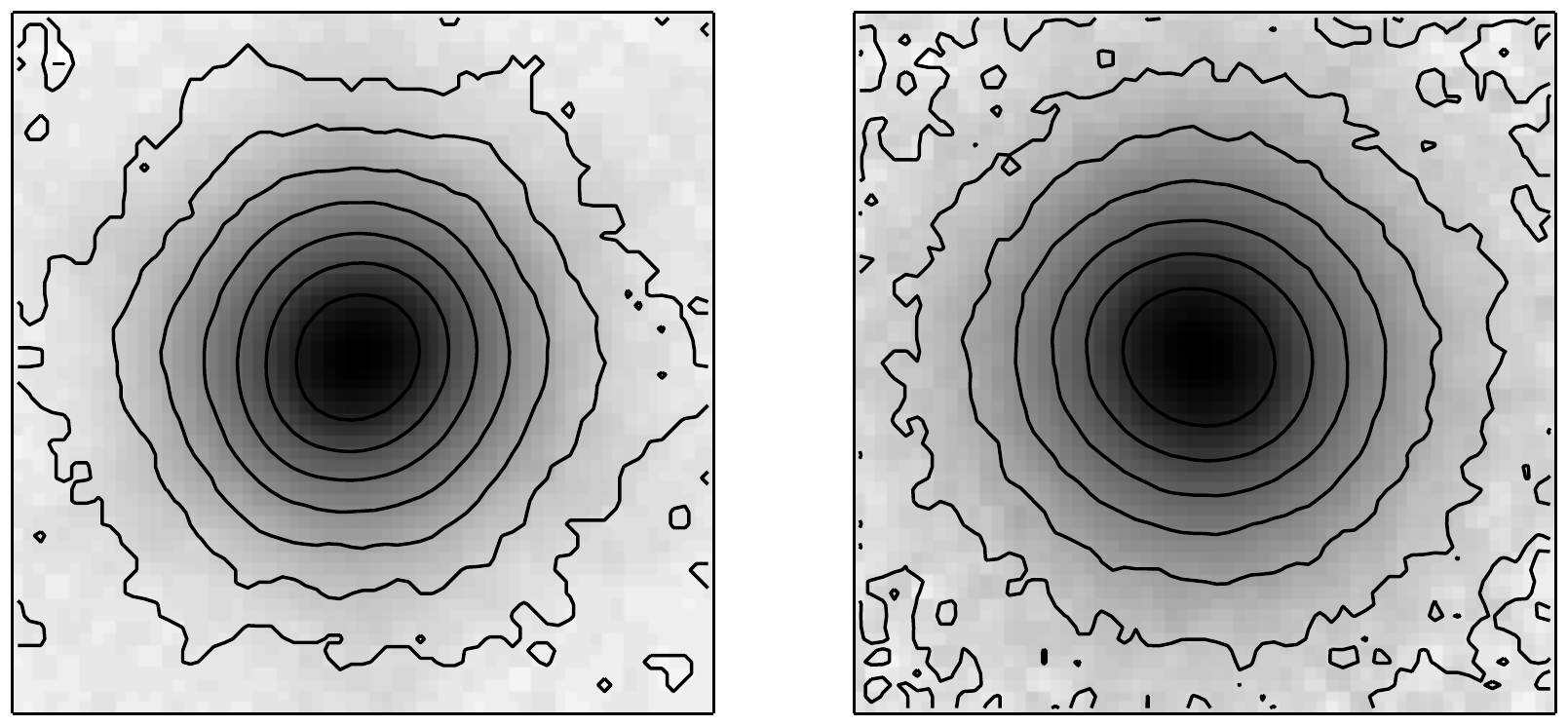}
\caption{The PSF of OSIRIS ({\it left}) and FORS2 ({\it right}) shown in logarithmic scale for images taken in similar seeing conditions. The OSIRIS image replicates the hexagonal  structure of the telescope pupil. Each panel measures $30\times30$ px or $\sim3\farcs8\times3\farcs8$.}
\label{psf_w}
\end{figure}

{\label{wings}}
Computation of photocentre positions in crowded fields or for close star pairs requires the iterative subtraction of the light from a neighbouring star image. This is realised by using a calibration PSF mask which {is a pixelised accumulation of the PSF profile and} represents well the actual shape of the PSF out to distances of 15 pixels from the centre \citep{Lazorenko:2014aa}. The calibration image  is compiled {for every frame} by accumulating the pixel counts in images of bright isolated stars after normalizing to some standard star brightness.  Figure \ref{psf_w} illustrates the calibration PSFs for two single images taken with OSIRIS and FORS2 obtained at comparable seeing with FWHM=0\farcs71. For both cameras, the PSF shape is very similar, except that the outer wings of the OSIRIS image display the symmetric hexagonal  pattern that reflects the shape of the primary mirror. This feature is weak and well modeled, and therefore should not degrade the astrometry.  

\subsubsection{Average FWHM}{\label{fwhm_size}}
Both OSIRIS and FORS2 observations were acquired in service/queue mode and the distribution of FWHM therefore depends on a combination of requested seeing conditions, the program priority assigned by the time allocation committee, and the observatory science operations. For OSIRIS and FORS2, we had requested seeing $\leqslant0\farcs9$ and $\leqslant0\farcs8$, respectively.

\begin{figure}
\includegraphics[width=\linewidth]{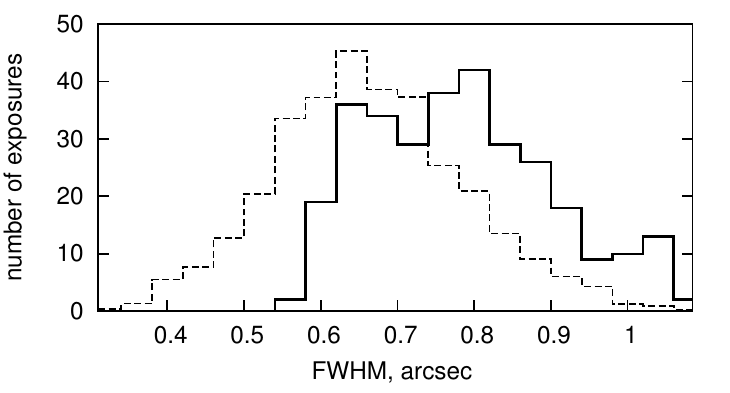}
\caption {Histogram of FWHM distribution for OSIRIS images (solid) and  FORS2  distribution normalised to the same number of exposures (dashed line). }
\label{hist}
\end{figure}

The distribution of FWHM in all used OSIRIS frames\footnote{This sample does not include the  frame series excluded because of poor image quality or telescope pointing.} is shown in Fig. \ref{hist} and compared to the results of the FORS2 data \citep{Lazorenko:2014aa} scaled to equal number of exposures. The FWHM of OSIRIS images is 0\farcs80 on average, which is $\sim$23\% larger than the corresponding mean FWHM of $0\farcs65$ for FORS2. The uncertainty of the photocentre positions is larger for OSIRIS by the same extent, because it is proportional to FWHM. 

\begin{figure}
\includegraphics[width=\linewidth]{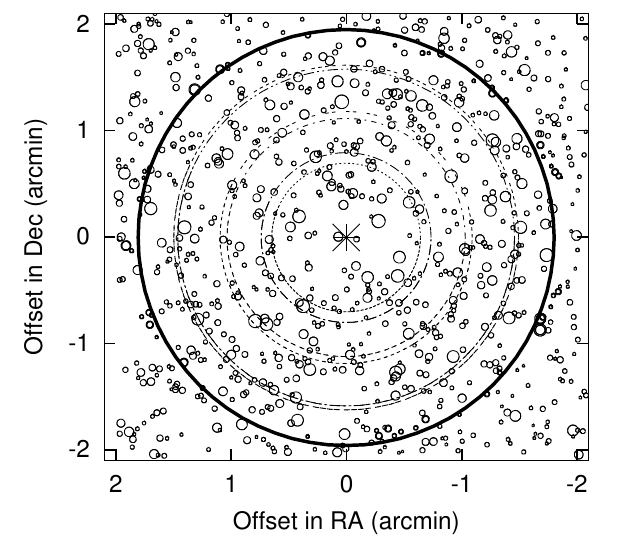}
\caption{Positions of stars on the CCD2 chip of OSIRIS that were used for the astrometric reduction. \dw\ is marked with an asterisk and the sizes of open circles indicate the brightnesses of reference stars. All stars are located within 2\arcmin\ of the target and six concentric circles with increasing radii mark the reference fields used for the reduction with $k=6 -16$ (Sect. \ref{astr}). North is up and east is left. The outer bold circle delimits the stars included in the the catalogue (Sect. \ref{field}). The shown area corresponds to the green square in Fig. \ref{fov}.}  
\label{map}
\end{figure}

\subsection{Astrometric reduction}{\label{astr}}
The astrometric reduction is aimed at determining the position of the target in every image relative to its location in a reference frame defined by a standard image (or its equivalent based on several images). The  mapping of  the reference star positions between frames is performed using basic functions which are polynomials in $x$,$y$ of order $0,1, \ldots k/2-1$. The even integer $k$ is  called the mode of astrometric reduction. The model accounts for the reference frame distortion caused by the telescope optics and the image motion, and it takes into account the displacements of reference stars in time due to the proper motion,  parallax, and differential chromatic refraction. The system of astrometric parameters derived by this  method (proper motions, offsets, chromatic parameters, and parallaxes) is constrained by a set of conditions similar to the ones imposed on the residuals of a least-squares fit via basic functions  \citep{Lazorenko2009}. In particular, if the star parallaxes are randomly distributed in the FoV, the difference between the measured and true parallaxes  is a constant, usually referred to as the zero-point of parallaxes. For more complicated spatial distributions of parallaxes, this difference is not flat and slowly varies with $x$,$y$.

We used $k$ values in the range of 6--16 and the corresponding radii $R$ of reference star groups increased with $k$ from 0\farcm6 to 1\farcm48, leading to  number of reference stars between 56 and 361. The geometry of the reference fields is shown in Fig. \ref{map}. Some stars are located beyond the largest reference field of \dw, because we processed not only the target itself, but also a number of nearby stars in order to detect, estimate, and mitigate space-dependent systematic errors. The reduction of these stars is performed with their respective reference fields and stars in the outer parts of Fig. \ref{map} are also used as references to produce the catalogue presented in Sect. \ref{field}.

\subsubsection{Atmospheric image motion}{\label{sig1}} 
An important part of the astrometric error budget is the amplitude of the random differential image motion, which for a single frame, a fixed exposure time, and the mode $k$ is approximated by the model
\begin{equation}
\label{eq:atm}
\sigma_{\rm{at}}=B_k R^{b_k}
\end{equation}
The model parameters  $B_k $ and $b_k $  depend mainly on the atmospheric turbulence at the site \citep{Lazorenko2006}. For every target, they are adjusted to the values that lead to the best solution and the astrometric computations are performed in an iterative way starting from a set of initial values ($B_k^\prime $ and $b_k^\prime$) obtained from a default model \citep{Lazorenko:2014aa}. The optimal estimate can be expressed as $B_k=E_k B_k^\prime $, where the value of the correction factor $E_k$ indicates  how much  the actual image motion differs from the default model. In the case of FORS2,  the factor $E_k $ varies between 0.5 and 2.0 and is close to 1.5 on average. We applied  the same method for OSIRIS and found that the value of $E_k$ varies from 2.1 to 4.3 depending on $k$, with an average value of 3.3. This means that the atmospheric image motion for OSIRIS is about 2.2 larger compared to FORS2, which leads to large astrometric errors. More specifically, the value of $\sigma_{\rm{at}}$ which we measure is generated by the turbulence at high atmospheric altitudes, whereas the input from the lower turbulent layers is filtered out during the astrometric reduction \citep{Lazorenko2006}. We caution that with this model we cannot distinguish between atmospheric image motion and effects related to instability of the telescope and its optics. Therefore, the atmospheric image motion may be smaller in reality.

\subsubsection{Residuals of the astrometric model}
Using the transformation model, we obtain the astrometric measurements of the target  $\alpha^{\star}_m$\footnote{We use the notation $\alpha^{\star} = \alpha \cos{\delta}$ throughout the text.} and $\delta_m$ in RA and Dec, respectively, in frame $m$ at time $t_m$ relative to the reference frame of background stars and model them with six free parameters ($\Delta\alpha^{\star}_0, \Delta\delta_0, \mu_{\alpha^\star}, \mu_\delta, \varpi$, and $\rho$):
\begin{equation}\label{eq:axmodel}
\begin{array}{ll@{\hspace{2mm}}l}
\!\alpha^{\star}_{m} =& \Delta \alpha^{\star}_0 + \mu_{\alpha^\star} \, t_m + \varpi \, \Pi_{\alpha,m} &-\;\; \rho\, f_{1,x,m} \\
\delta_{m} = &{\Delta \delta_0 + \,\mu_\delta      \,  \;                      t_m \;+ \varpi \, \Pi_{\delta,m}}  &{+\;\; \rho \,f_{1,y,m}},
\end{array}
\end{equation}
where $\Delta\alpha^{\star}_0, \Delta\delta_0$ are coordinate offsets, $\mu_{\alpha^\star}, \mu_\delta$ are proper motions, and the parallactic motion is expressed as the product of relative parallax $\varpi$ and the parallax factors $\Pi_\alpha, \Pi_\delta$. As for the atmospheric refraction {modelled by $\rho$} in Eq. (\ref{eq:axmodel}), this model has one parameter less than the one used for our FORS2 work \citep{Lazorenko:2011lr, Sahlmann:2014aa} because the GTC does not incorporate a dispersion compensator. For OSIRIS, the differential chromatic refraction (DCR) is modelled with the free parameter $\rho$ and the quantity $f_{1}$, which is a function of  zenith angle, temperature, and pressure \citep{Lazorenko2006, Sahlmann:2013ab}.

To evaluate the quality of the astrometric reduction (and eventually to determine the astrometric parameters) we inspect the frame  residuals after adjusting the linear model Eq. (\ref{eq:axmodel}) to the data, which is accomplished by matrix inversion. By averaging the frame residuals of every epoch, we obtain the epoch residuals $x_{\rm ep}$ and $y_{\rm ep}$ in RA and Dec, respectively.

This reduction technique is applicable equally both to \dw\ and any field star with its unique set of reference stars. Astrometric solutions for field stars are useful to test the presence of systematic errors, to derive the parallaxes  of background stars, for the correction to absolute parallax, to determine the pixel scale, and for compiling the catalogue of field stars. 

\subsubsection{Non-standard observations}{\label{non_standard}
The default position of the target was chosen to be near the CCD2 centre. In the images of epoch No. 7, however, the  star field was displaced by $dx \approx 90$~px along the $x$-axis. For epoch No. 9, a similar offset occurred along both axis and the images were slightly asymmetric, although with good FWHM=0\farcs8. To investigate the effect of non-standard telescope pointing on the astrometry, we ran a test reduction  for  30 bright stars near the field centre.  Each of these test star was processed in the same way as the target, with the epoch residuals $x_{\rm ep}$, $y_{\rm ep}$ as output data. We used the images of these two problem epochs and also included the images of epoch No. 1, which exhibit clearly non-symmetric PSF shapes with multiple peaks and were obtained with FWHM=0\farcs93, i.e.\ slightly above the average seeing. Usually, such images are discarded during the preliminary inspection. The computations for the total of 14 epochs were run with $k=10$ only and  Figure \ref{ep_disp} shows the scatter of the epoch residuals for these stars, which is abnormally large for the epochs No. 1, 7, and 9.  

\begin{figure}
\includegraphics[width=\linewidth]{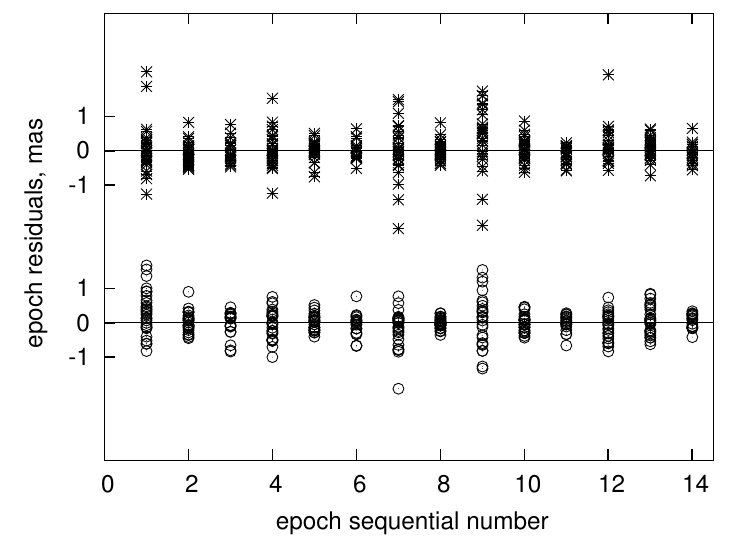}
\caption {The  epoch residuals $x_{\rm ep}$ (open circles) and $y_{\rm ep}$ (asterisks) for 30 bright stars as a function of epoch number. There is abnormally large scatter for epochs No. 1, 7, and 9.}
\label{ep_disp}
\end{figure}

The large scatter for the epoch No.1 residuals is expected, because the important distortion of the PSF shape leads to effects which cannot be adequately compensated with our reduction method, thus eventually creates systematic errors in differential positions. Although the FWHM of the images in epochs No.1 and No. 9 are within the tail  of the distribution shown in Fig.\,\ref{hist}, the scatter of their epoch residuals $\sigma_{\rm ep}$ is increased approximately two-fold, with many deviations larger than 1 mas. 

The field offsets in epoch No. 7 and No. 9 have a different effect:  when the geometry of the CCD is not perfect, for instance if the CCD pixel columns are curved, the field offset can affect the epoch residuals $x_{\rm ep}$, $y_{\rm ep}$ because the local segments of CCD columns for the target and field stars can be relatively inclined, say at some angle $i$. If the field offset is $dx$, $dy$ along the CCD axes, it results in the displacement of the CCD positions on $i dy$ and $-i dx$ along RA and Dec, respectively. Because the value of $i$ depends on the star position in the FoV, the effect is different in RA and Dec in terms of amplitude and sign. As a result, we expect excess random scatter of the epoch residuals in RA for the epoch No.9, and in Decl for the epochs No. 7 and No. 9, which is confirmed in Fig.\,\ref{ep_disp}.

\subsubsection{Epoch residuals for nominal observations}{\label{ep_res}}
We repeated the astrometric reduction with the eleven nominal epochs and computed the r.m.s. of the measured single frame residuals in $x$ and $y$. Both for \dw\ and for a sample of bright field stars, each reduced as a target, we obtained a value of $1.05$~mas. This matches the expected single-frame precision obtained from a model that accounts for the precision of the photocentre determination, the atmospheric image motion, and the reference field noise. Hence, the expected nominal precision of the epoch residuals $x_{\rm ep}$, $y_{\rm ep}$ is  $\sigma _{\rm nom}=0.20$~mas for a typical epoch consisting of 28 frames. This is a lower limit, which does not take into account systematic errors that affect the epoch positions but are constant during the epoch. The more complete astrometric reduction described in \citet{Lazorenko:2014aa} allowed us to uncover the existence of systematic errors, of which some are correlated in CCD space. Their contribution to the error budget slightly increases the expected nominal epoch precision to $\sigma _{\rm nom}=0.23$ mas. %

We compare this value to the r.m.s. of the epoch residuals $x_{\rm ep}$, $y_{\rm ep}$, which is $\sigma_{\rm ep}=0.40$~mas and $\sigma_{\rm ep}=0.35$~mas for \dw\ and the bright stars, respectively. Those two values are comparable, meaning that the red and fast-moving target is measured with the same precision as field stars and it excludes the presence of orbiting companions that would introduce a measurable astrometric signature. However, the difference between the measured epoch precision of 0.35--0.40~mas and its expected value of 0.23~mas is significant and points towards unknown systematic errors.

\subsubsection{Relation between $\sigma_{\rm ep}$ and $\sigma(\psi)$}
Finally, we ascertained that there is no correlation between $\sigma_{\rm ep}$ and $\sigma(\psi)$ for bright field stars. This means that if images are compact and visually symmetric,  implying that the degree of deformations of the PSF kernel is $\sigma(\psi) \leq 0.03$, they  yield essentially the same results as ideal images with $\sigma(\psi)=0$, i.e.\ with no excess of systematic errors in the differential positions.

\section{Results}
To obtain the results of our astrometric study of \dw, we followed mostly the prescriptions of \cite{Sahlmann:2014aa}.

\subsection{Astrometric parameters of \dw}\label{sec:astroparam}
We obtained the astrometric parameters of our target \dw\ by obtaining the least-squares solution of Eq.~(\ref{eq:axmodel}) for the photocentre positions determined from the OSIRIS images. The solution was found using matrix-inversion and takes into account the measurement uncertainties and covariances. The parameters are given in Table\,\ref{par_tbl} and Figs.\,\ref{sky_plot} and \ref{time_plot} illustrate the results. The reference date $T_\mathrm{Ref}$ was chosen as the mean observation date to minimise parameter correlations, which however do not vanish because of the actual sampling of our observations, see Table \ref{tab:dwcorr}.

\begin{table}
\caption{Astrometric parameters of \dw. Standard uncertainties were computed from the parameter variances that correspond to the diagonal of the problem's inverse matrix and rescaled to take into account the residual dispersion.}
\centering
\begin{tabular}{lcr}
\hline
$\Delta \alpha^\star$ & (mas) & $27.84 \pm 0.09$           \\ 
$\Delta \delta      $  & (mas) & $-190.36 \pm 0.12$             \\ 
$\varpi              $  & (mas) & $106.15 \pm 0.18$   \\ 
$\mu_{\alpha^\star}$  & (mas yr$^{-1}$) & $230.27 \pm 0.16$   \\ 
$\mu_{\delta}$         & (mas yr$^{-1}$) & $-241.49 \pm 0.12$   \\ 
$\rho$                  & (mas) & $22.08 \pm 0.13$              \\ 
\multicolumn{3}{c}{Derived and additional parameters}\\[3pt]
$T_\mathrm{Ref}$ &(MJD)& {56639.124721}\\[1pt]
$\Delta\varpi$& (mas)&    $-0.50\pm0.05$ \\
$\varpi_\mathrm{abs}$ & (mas)&     $106.65\pm 0.20$ \\
Distance & (pc)&     $9.38\pm 0.03$\\
\multicolumn{2}{l}{Number of epochs / frames}  & { 11 / 312}\\
$\sigma_\mathrm{ep}$ & (mas)  & {0.401}\\ 
\hline
\end{tabular}
\label{par_tbl}
\end{table}

\begin{table} 
 \caption{Parameter correlation matrix.}
  \centering 
\begin{tabular}{l |r@{\;\;} r@{\;\;} r@{\;\;} r@{\;\;} r@{\;\;} r@{\;\;} r@{\;\;}} 
	\hline\hline 
&$\Delta \alpha^\star_0$ &$\Delta \delta_0$ &$\varpi$ &$\mu_{\alpha^\star}$ &$\mu_{\delta}$ &$\rho$ \\ \hline
$\Delta \alpha^\star_0$ &$+1.00$ & \\ 
$\Delta \delta_0$ &$-0.04$ &$+1.00$ & \\ 
$\varpi$ &$+0.44$ &$-0.77$ &$+1.00$ & \\ 
$\mu_{\alpha^\star}$ &$+0.41$ &$-0.49$ &$+0.69$ &$+1.00$ & \\ 
$\mu_{\delta}$ &$+0.20$ &$-0.28$ &$+0.43$ &$+0.48$ &$+1.00$ & \\ 
$\rho$ &$+0.39$ &$+0.56$ &$-0.27$ &$-0.11$ &$-0.13$ &$+1.00$   \\ 
\hline 
 \end{tabular} 
 \label{tab:dwcorr}
\end{table} 

\begin{figure}
\includegraphics[width=\linewidth]{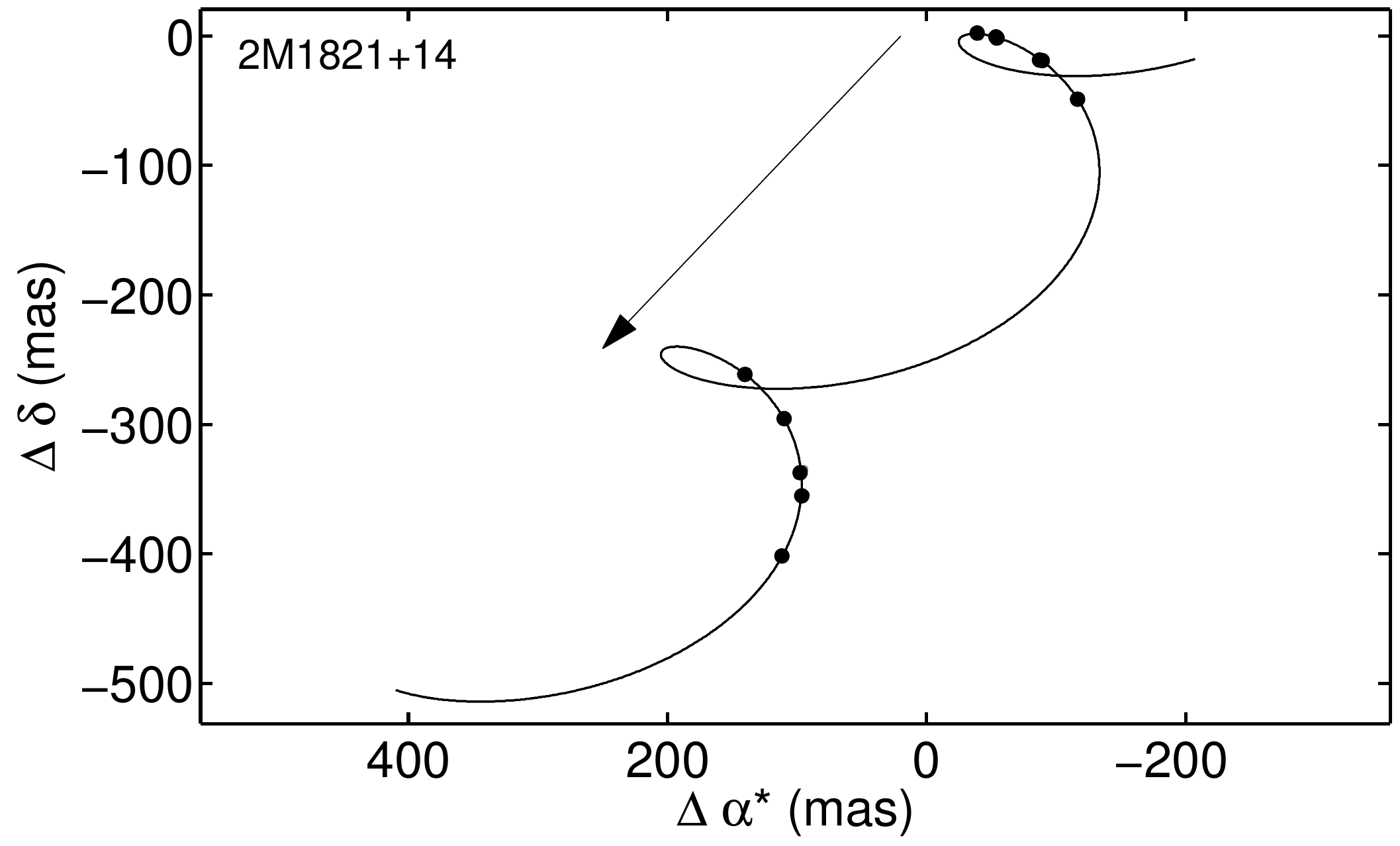}
\caption {OSIRIS astrometric measurements of \dw\ (circles) and the best-fit model that includes parallax and proper motion (solid curve). The arrow indicates the proper motion direction and amplitude in one year. The epoch uncertainties are comparable or smaller than the symbol size. North is up and east is left. }
\label{sky_plot}
\end{figure}
\begin{figure}
\includegraphics[width=\linewidth, trim= 0 0 5cm 1.1cm, clip=true]{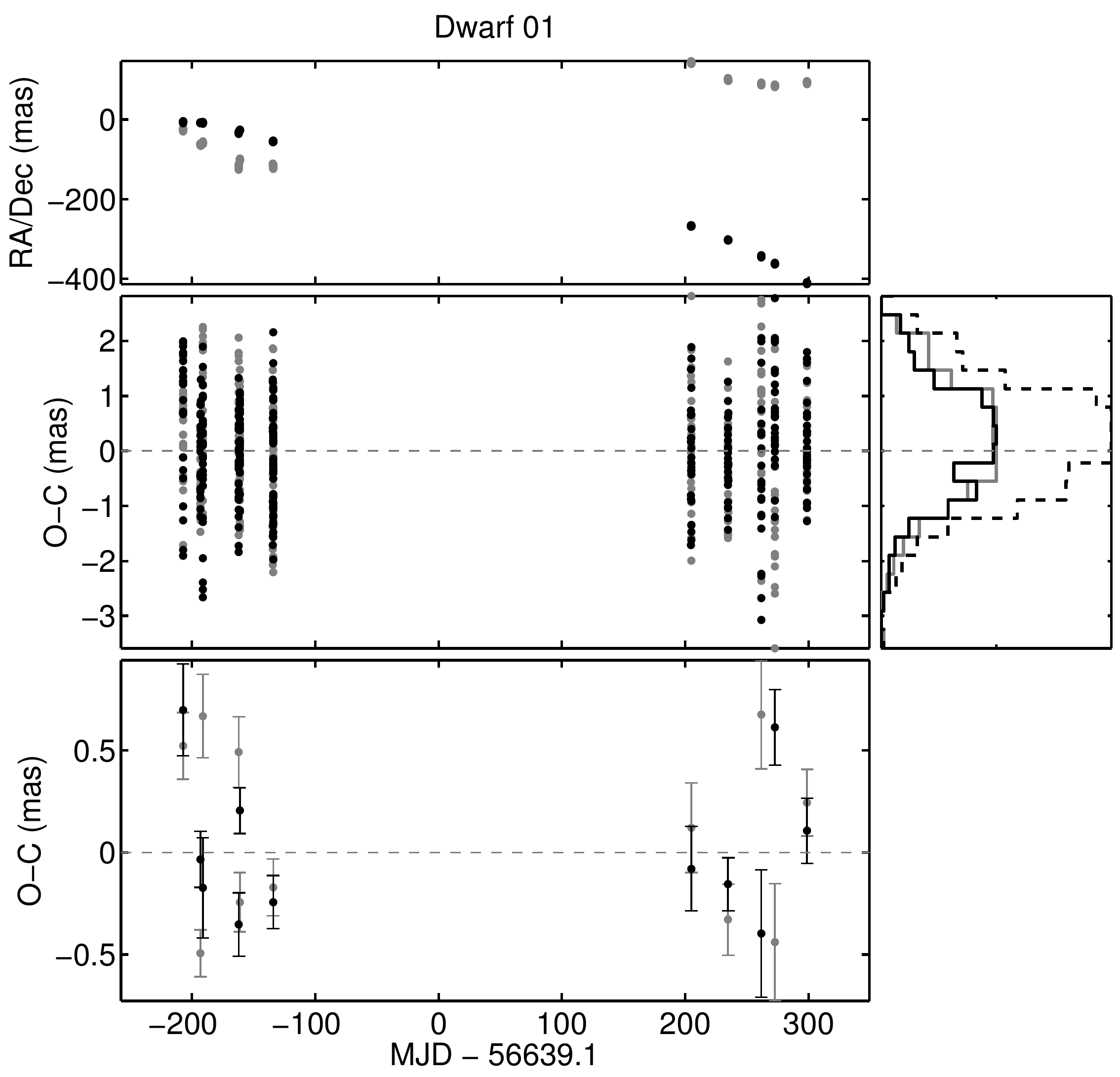}
\caption {Residuals of the astrometry solution of \dw\ as a function of time. The top panel shows the reduced data in RA (grey symbols) and Dec (black symbols) and the middle panel shows the frame residuals after solving Eq. (\ref{eq:axmodel}). The bottom panel shows the epoch-averaged residuals with their mean uncertainties.}
\label{time_plot}
\end{figure}

Because the astrometric reference stars are not located at infinity, a correction has to be applied to the relative parallax to convert it into an absolute parallax that can yield the distance. As in \cite{Sahlmann:2014aa}, we use the the Galaxy model of \cite{Robin:2003fk} to obtain a large sample of pseudo-stars in the region around \dw. The comparison between the model parallaxes and the measured relative parallaxes of stars covering the same magnitude range yields an average offset, which is the parallax correction. Using $N_\mathrm{stars}=450$ reference stars, we obtained a parallax correction of $\Delta \varpi_\mathrm{galax}=-0.50$ mas with an r.m.s. value of $\sigma_\mathrm{galax} = 1.10$ mas. The absolute parallax is $\varpi_{abs} = \varpi - \Delta \varpi_\mathrm{galax}$ and its uncertainty was computed by adding $\sigma_\mathrm{galax}/\sqrt{N_\mathrm{stars}}$ in quadrature to the relative parallax uncertainty. Figure \ref{fig:7} shows the magnitudes on pseudo- and reference stars, demonstrating a good match of their distributions. 

In principle, a similar procedure should be applied to correct from relative to absolute proper motion. We refrain from doing so, because proper motion is a less critical parameter in the following analyses and the correction will be small. Our proper motions agree with the values derived by \citealt{Gagne:2014aa}, however we determined them with $\sim$50 times smaller uncertainties. In the future, the results of ESA's \emph{Gaia} mission will make it possible to determine model-independent parallax and proper motion corrections, because \emph{Gaia} will obtain accurate astrometry for many of the reference stars used here. 

\begin{figure}
\centering
\includegraphics[width=0.9\linewidth]{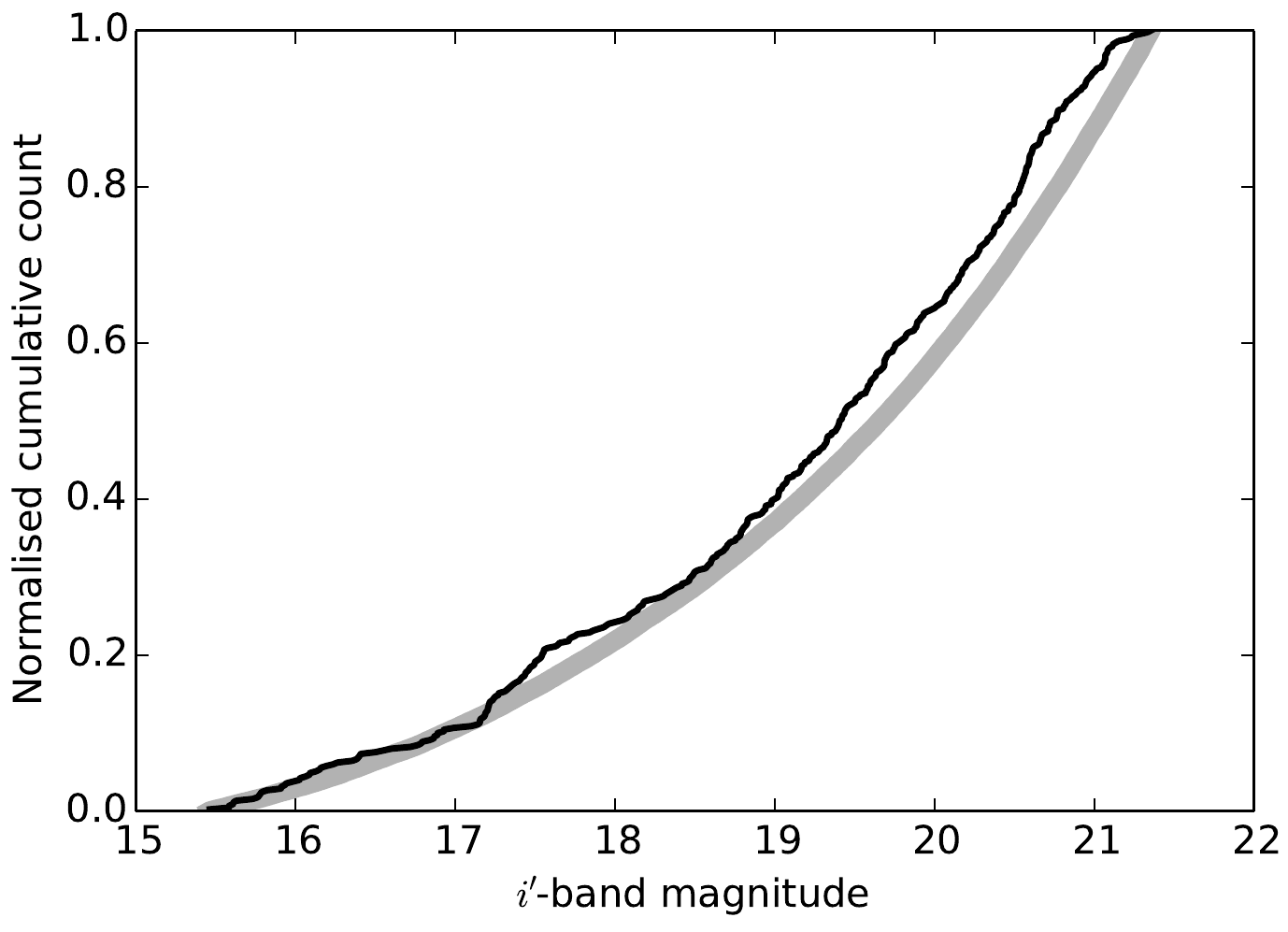}
\caption{Using a Galaxy model to determine the parallax correction. Cumulative distribution of magnitudes for the used reference stars of \dw. The model and measured data are shown in grey and black, respectively.}
\label{fig:7}
\end{figure}

The resulting distance of $9.38 \pm0.03$ pc agrees with the spectrophotometric estimate of $\sim$10 pc of \cite{Looper:2008aa}. Figure \ref{fig:HRD} shows the $K$-band absolute magnitude of \dw, which appears comparable to other field objects with measured precision parallaxes. This suggest that the metallicity of \dw\ does not differ significantly from that of the field, and we will therefore assume that is has solar metallicity.

\begin{figure}
\centering
\includegraphics[width=\linewidth]{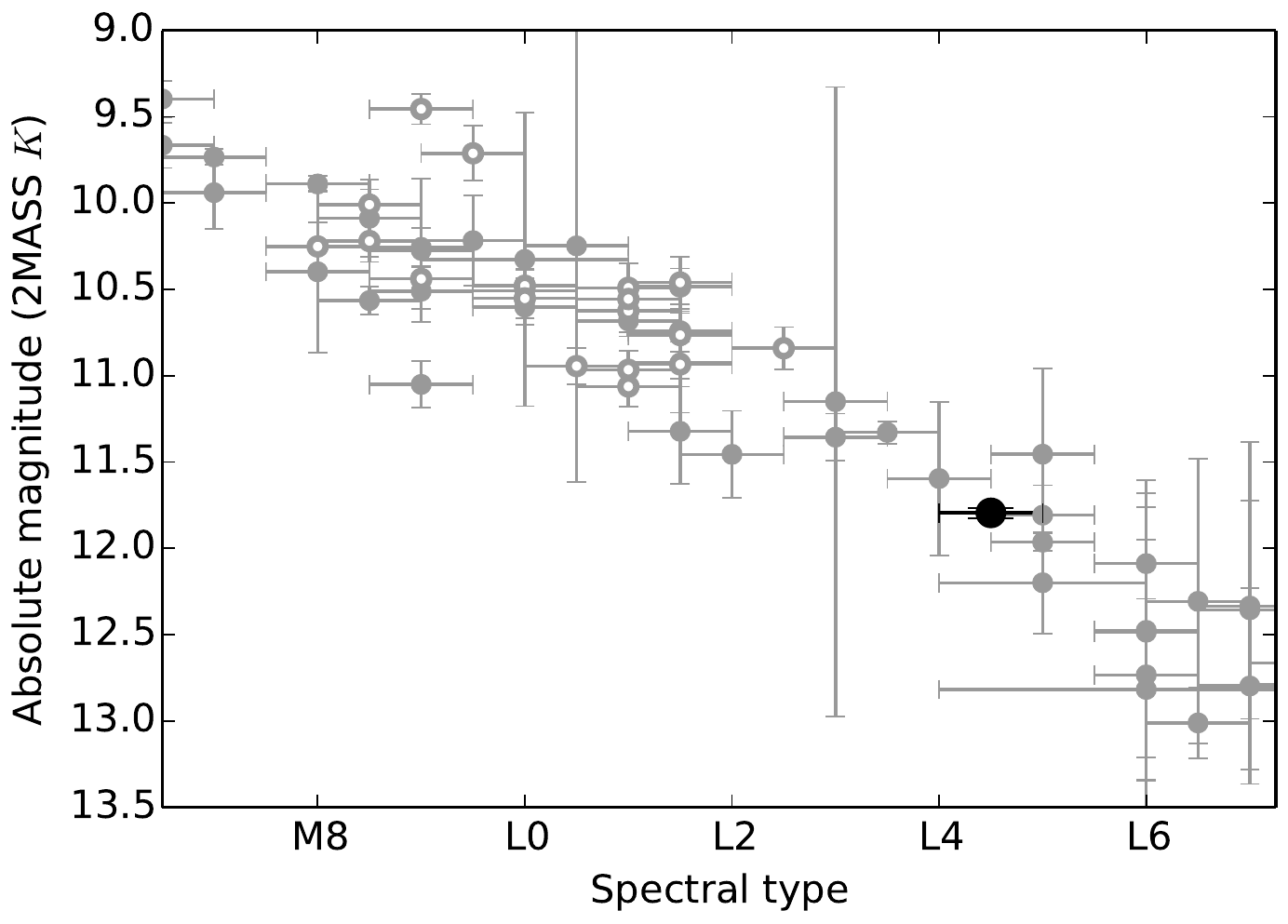}
\caption{Absolute magnitude in the 2MASS $K$-band as a function of spectral type for M6--L7 dwarfs in the database of ultracool parallaxes {\citep[filled grey  symbols,][]{Dupuy:2012fk}}, for the FORS2 survey sample {\citep[open grey symbols,][]{Sahlmann:2014aa}}, and for \dw\ with a spectral type of L4.5 (black circle).}
\label{fig:HRD}
\end{figure}

\subsection{Estimating the age of \dw}
\dw\ was identified as a potentially young member of the field population that shows signs of low surface gravity \citep{Looper:2008aa, Gagne:2014aa, Yang:2015aa}. The mid-L spectral type combined with the clearly detected lithium absorption at 670.8 nm in the optical spectrum \citep{Looper:2008aa} implies that its mass has to be smaller than about 0.06 $M_{\odot}$ \citep{Magazzu:1993kx}. We thus assume a preliminary upper limit of $\sim$1 Gyr for the age of \dw, which we refine in the following.

Using the radial velocity of 9.8 $\pm$ 0.16 km\,s$^{-1}$ \citep{Blake} and the proper motions and absolute parallax of this paper, we derived the $U$, $V$, and $W$ heliocentric velocity components in the directions of the Galactic center, Galactic rotation, and north Galactic pole, respectively, with the formulation provided by \cite{Johnson:1987aa}. We used the right-handed system and the solar motion is not subtracted from our calculations. The uncertainties associated with each space velocity component are obtained from the quoted parallax, proper motion (with 1 mas\,yr$^{-1}$ uncertainties to account for the unknown but small offset to their absolute values), and radial velocity uncertainties after the prescriptions of \cite{Johnson:1987aa}. For \dw\ we find $U$ = 12.91 $\pm$ 0.12, $V$ = 4.62 $\pm$ 0.11, and $W$ = $-11.30 \pm 0.06$ km\,s$^{-1}$. These values indicate a young kinematic age, which agrees with the results by \cite{Gagne:2014aa}. According to \cite{Eggen:1990aa} and \cite{Leggett:1992fk}, these Galactic velocities are typical of the young-old disk. Although the Galactic velocities of \dw\ do not fall within the 1-$\sigma$ ellipsoids of known young stellar moving groups \citep[e.g.][]{Zuckerman:2004fk, Torres:2008aa, Zuckerman:2011aa} in $UVW$ diagrams, they are relatively close to the Ursa Major moving group (see Fig. \ref{fig:uvw}), which has an estimated age of 300--500 Myr. 

With the spectroscopic rotational velocity of $v \sin i = 28.85 \pm 0.16$ km\,s$^{-1}$ determined by \cite{Blake} and the photometric periodicity of $4.2 \pm 0.1$ h measured by \cite{Metchev:2015aa} that is ascribed to rotation, we can calculate the minimum radius of $R \sin i = 0.100 \pm 0.003 \,R_{\odot}$. This implies that the true radius of \dw\ is 0.10 $R_{\odot}$ or larger. By comparison with the Lyon group's theoretical evolutionary substellar models, it is also inferred that the age of this source is younger than 1 Gyr, see Fig. \ref{rot}. For an angle of the rotation axis near 90\degr, the age of \dw\ would be close to 700 Myr. 

A tentative lower age limit can be derived from comparison with the location of Pleiades ($\sim$120 Myr) and field objects in color-magnitude diagrams, of which one is shown in Fig. \ref{fig:age}.  \dw\ appears closer to the field than to the Pleiades and it is not as red and and luminous as Pleiades members of similar spectral type. Therefore, the most likely age of \dw\ lies in the range of 120--700 Myr and we adopt a value of 500 Myr.

\begin{figure}
\includegraphics[width=\linewidth,trim=1cm 0.5cm 2cm 9.5cm,clip]{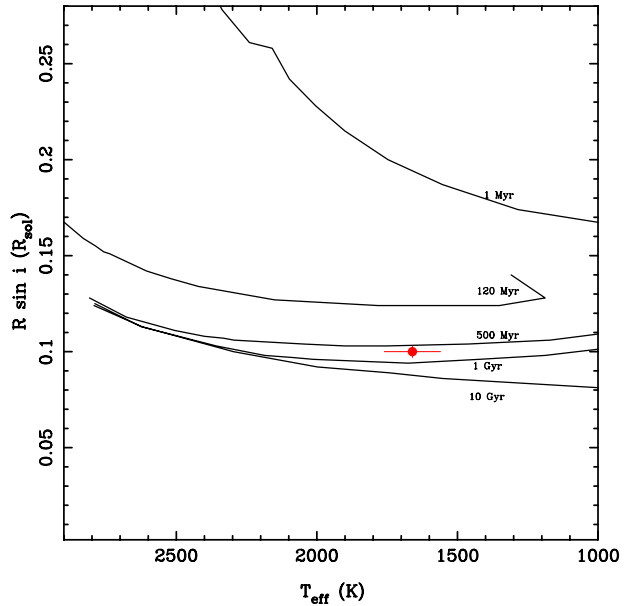}
\caption{Radius as a function of $T_\mathrm{eff}$ for \dw\ (red circle) using the \citet{Stephens:2009aa} $T_\mathrm{eff}$ -- spectral type relationship. The isochrones shown as solid lines correspond to the models of \citet{Chabrier:2000kl}.}
\label{rot}
\end{figure}

\begin{figure}
\includegraphics[width=\linewidth,trim=1cm 1cm 2cm 9cm,clip]{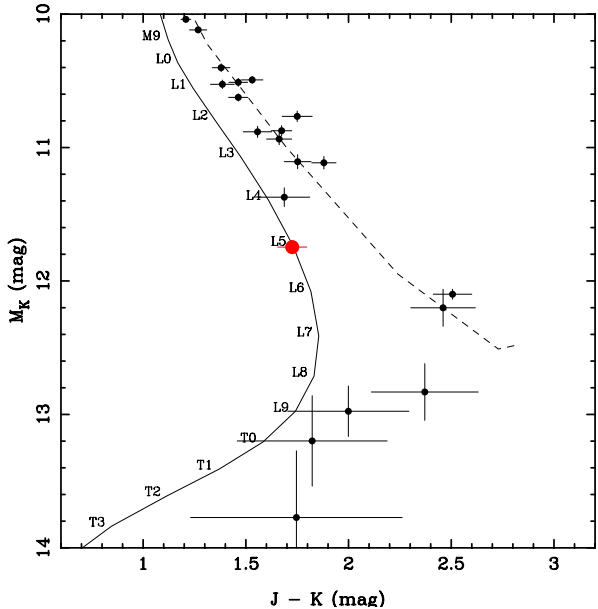}
\caption{Colour-magnitude diagram showing \dw\ (red large circle) and Pleiades members (black small dots) from \citet{Bihain:2010kx} and \citet{Zapatero-Osorio:2014ac}. The solid line shows the field \citep[according to][]{Stephens:2009aa} and the dashed line represents the 120-Myr isochrone following \citet{Zapatero-Osorio:2014ac}.}
\label{fig:age}
\end{figure}

\subsection{Mass estimate of \dw}\label{sec:mass} 
To continue interpreting our results, in particular to derive limits on the masses of potential companions, a mass estimate of \dw\ is required. Because we rely on photometry, the estimated masses of ultracool dwarfs are tightly linked to their ages. Assuming an age range of 120--700 Myr, we used a method that relies on an estimation of the bolometric luminosity. We converted {\small 2MASS} magnitudes to the {\small MKO} system using updated colour transformations \citep{Carpenter:2001ys}\footnote{\url{http://www.astro.caltech.edu/~jmc/2mass/v3/transformations/}} and bolometric corrections \citep{Liu:2010fk} to obtain the luminosity. The corresponding mass at a given age was found by interpolating the DUSTY models \citep{Chabrier:2000kx}. Differences between $J$,$H$,$K$-bands are negligible, and we used their average. The resulting masses are listed in Table \ref{tab:masses} and include the uncertainties in apparent magnitude, parallax, and spectral type (assumed to be L4.5$\pm$1). Those formal uncertainties are sometimes smaller than the model uncertainties, which we assume to be 10 \% on the derived masses.

We see that for ages younger than 1 Gyr, \dw\ has a mass smaller than the lithium-burning mass limit, which is consistent with the presence of Li I absorption in its optical spectrum  \citep{Looper:2008aa}. According to this method the mass of \dw\ would be $0.049^{+0.014}_{-0.025}\,M_{\odot}$ for an age of $500^{+200}_{-380}$ Myr.

\begin{table}
\caption{Mass estimates for \dw\ with theoretical formal uncertainties.}
\centering
\begin{tabular}{r c}
\hline
Age & Mass \\
(Gyr) & ($M_{\odot}$) \\
\hline
0.12 & $ 0.024 \pm 0.004 $\\
0.50 & $ 0.049 \pm 0.001 $\\
1.00 & $ 0.063 \pm 0.001 $\\
\hline
\end{tabular}
\label{tab:masses}
\end{table}

\begin{figure*}
\includegraphics[width=\linewidth,trim=0 10.5cm 0 8.5cm,clip]{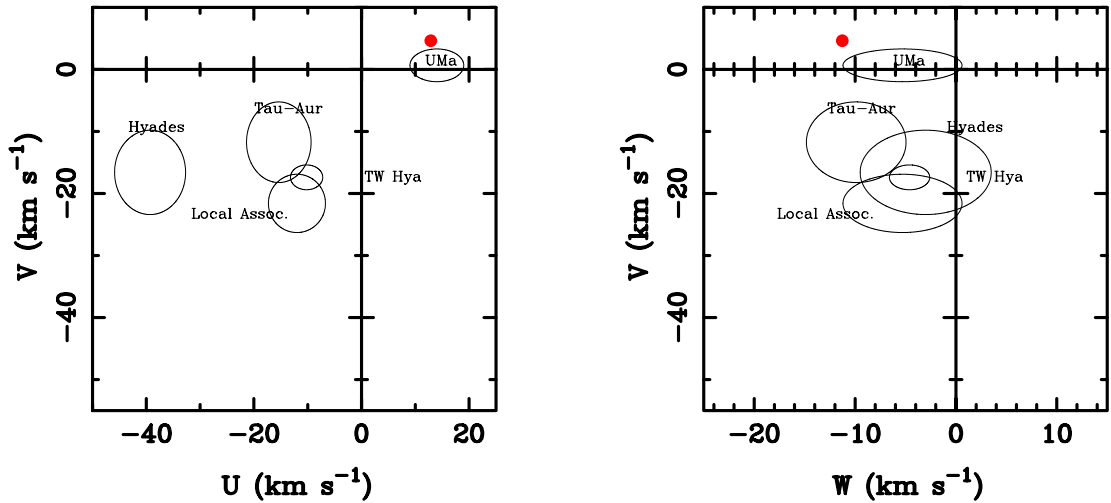}
\caption{$UVW$ velocities of \dw\ (red dot) and the 1-$\sigma$ ellipsoids of different young moving groups. The uncertainties for \dw\ have a size similar to the symbol.}
\label{fig:uvw}
\end{figure*}

\subsection{Planet detection limits}
One of the goals of this study is to evaluate the potential of OSIRIS astrometry for planet discovery. Already with the data collected here, we are able to exclude the presence of companions to \dw\ that have orbital periods comparable to the timespan of OSIRIS observations, i.e.\ $\sim$500 d. We estimated the companion detection limits following the procedure described in \cite{Sahlmann:2014aa}. This procedure assumes that the light contribution of the companion to the photocentre motion is negligible, which may not be the case for young and high mass-ratio systems \citep[cf.][]{Sahlmann:2015ab}. 

Figure \ref{fig:detlim} shows the maximum mass of any putative companion to \dw\ as a function of its orbital period and projected physical separation. In particular, for orbital periods between $\sim$50 and $\sim$1000 days ($\sim$0.1--0.7 au), the OSIRIS/GTC data exclude at 3-$\sigma$ confidence the presence of any companion down to a mass ratio of 0.1, corresponding to a planet with a mass of about 5 Jupiter masses ($M_\mathrm{J}$). The low mass and nearby distance of \dw\ are favourable to push the detection limit to small companion masses, despite the residual dispersion of 0.4 mas that is larger than the average value of 0.12 mas achieved for FORS2. 

\begin{figure}
\includegraphics[width=\linewidth]{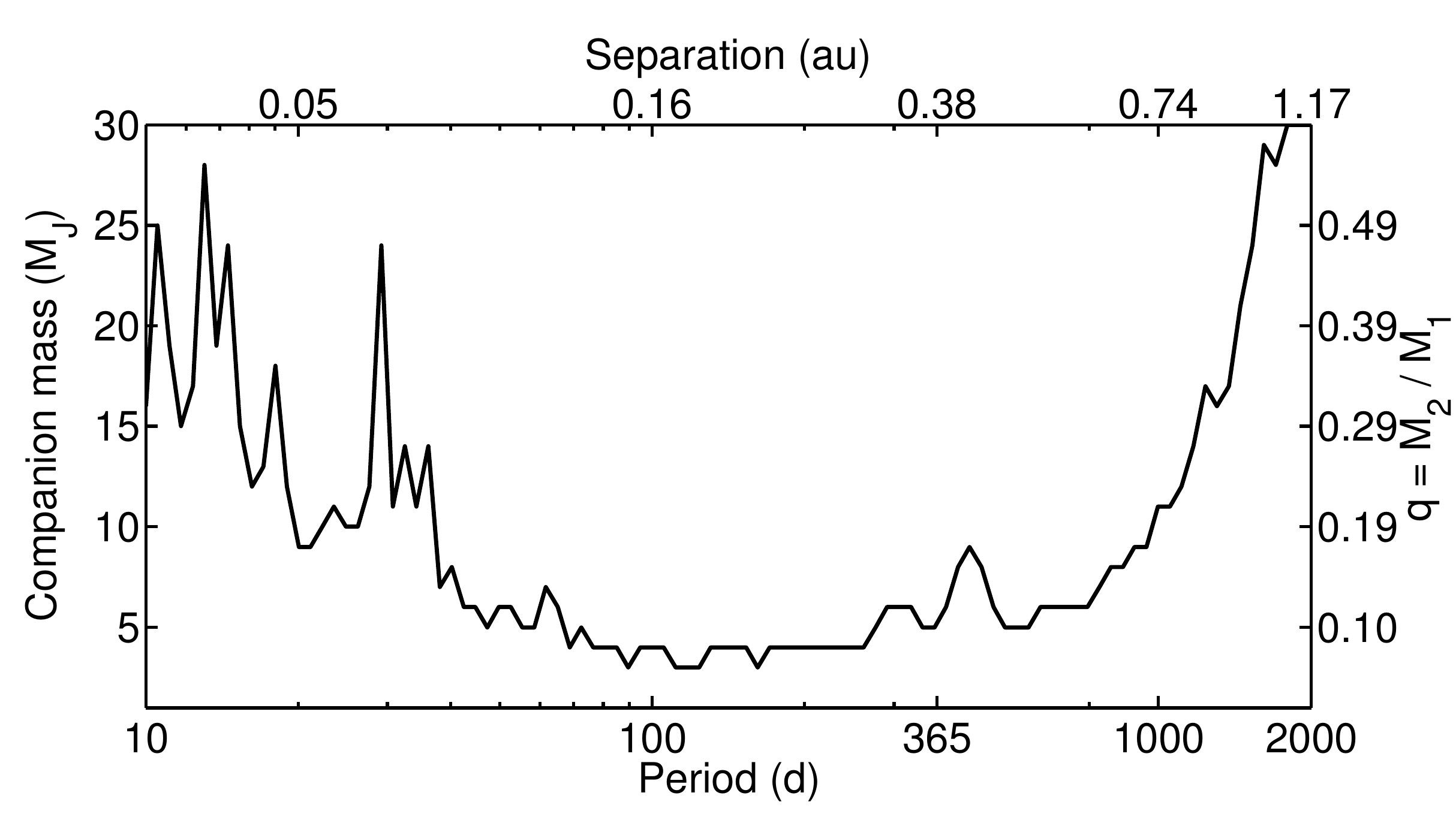}
\caption{Companion detection limits for \dw\ as a function of orbital period (bottom label) and relative primary-secondary separation (top label, computed for a 5 $M_J$ companion around a 0.049 $M_{\odot}$ primary). The right label indicates the mass ratio $q$. The maximum companion mass compatible with the measurements is shown, i.e.\ companions on and above the curve are excluded by the data at 3-$\sigma$ confidence.}
\label{fig:detlim}
\end{figure}

\subsection{Optical variability of \dw}
We performed relative photometry of \dw\ with the OSIRIS data using the method of \cite{Sahlmann:2014aa}, which is based on brightness measurements of \dw\ and dozens of field stars in each of the eleven astrometric epochs.  We used different sets of comparison stars (between 10 and 60)  and obtained very similar results. The photometric variation of \dw\ is 12 mmag r.m.s. in the Sloan $i'$-band as shown in Fig. \ref{fig:variability}. This is higher than the typical variability for reference stars that have values of 1--5 mmag with an average of 3 mmag r.m.s., a value that represents the uncertainty of the method. We used the reference star photometry to investigate correlations with airmass (which varied between 1.0 and 2.0) or sky conditions but found none. Our observations therefore suggest that \dw\ shows optical variability at $>$10 mmag level on timescales of weeks to years.

\dw\ was also monitored by \cite{Koen:2013uq}, who did not detect significant optical variability, and a near-infrared spectral variability study was performed by \citet{Yang:2015aa}. Finally, \cite{Metchev:2015aa} detected infrared variability with a period of $4.2\pm0.1$ h. We examined our OSIRIS photometry for periodic variations at similar timescales using a periodogram analysis, but did not find a significant signal. This may be explained by an amplitude that is much smaller in the optical than at 3.6/4.5\,$\mu$m or by the sampling of our data, which consists of eleven epochs with $\sim$55 min duration each, separated by several days or weeks. In addition, the 4.2 h {photometric} signal detected in the 10 h continuous timeseries of \cite{Metchev:2015aa} may not be {phase-}coherent over timescales of years, reducing the probability of detecting it with our OSIRIS photometry.

\begin{figure}
\center
\includegraphics[width= \linewidth]{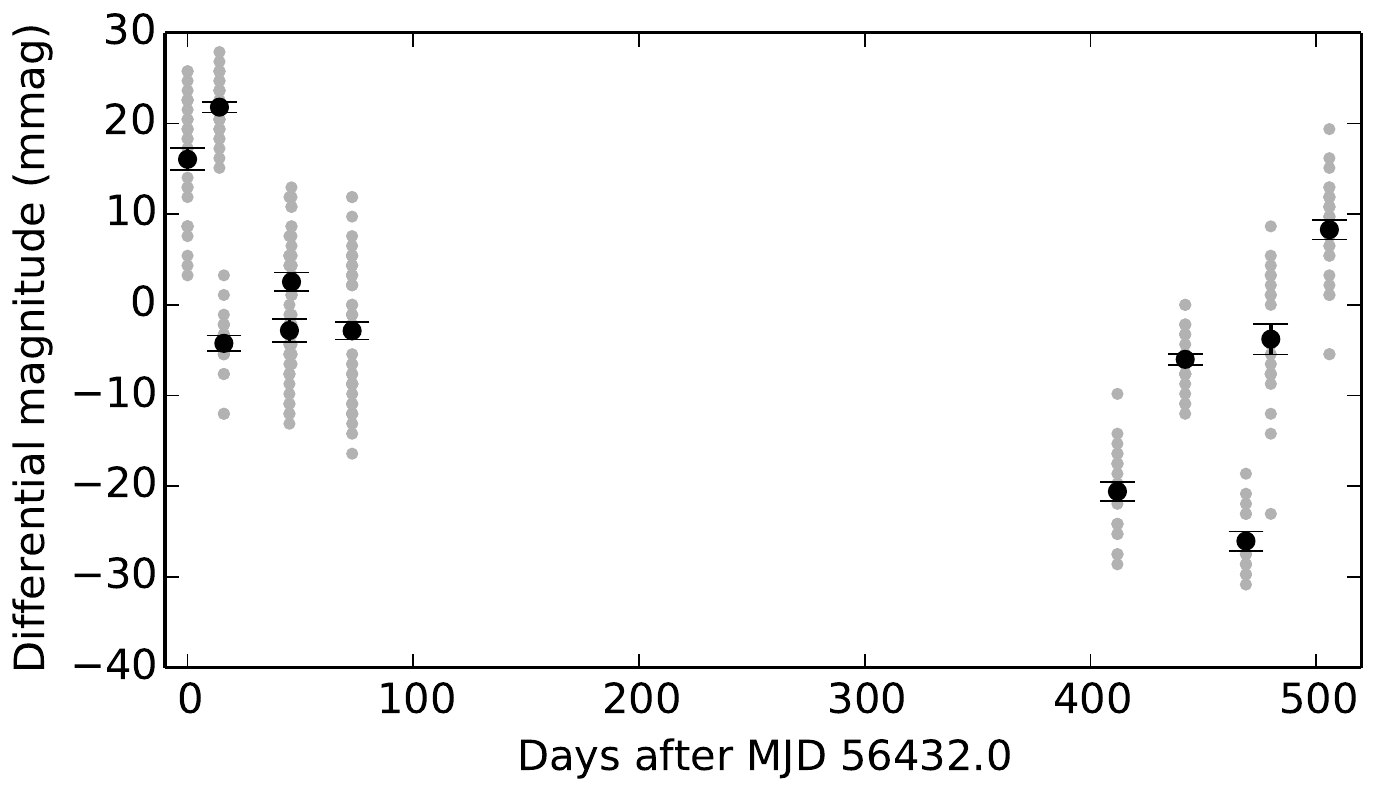}
\caption{Differential magnitude variation of \dw\ in $i'$-band as a function of time. Grey symbols correspond to measurements in individual OSIRIS frames, whereas black circles show the epoch averages with uncertainties given as error of the mean.}
\label{fig:variability}
\end{figure}

\subsection{Catalogue of field stars}{\label{field}}
We supplement this study with the astrometric parameters of the 587 stars with magnitudes $i'=15.5-22$ that were used as astrometric references in the field of \dw, i.e.\ that are located within $2\arcmin$ of the target. We computed their parameters with the reduction mode $k=10$. The typical precision of their parallaxes varies from 0.15 mas to 2.5 mas depending on brightness and the uncertainty of proper motions ranges from 0.1 mas/yr to 3 mas/yr. There are no stars with proper motions larger than $\pm 10$ mas/yr. 

Fig. \ref{parall} presents the {absolute parallaxes $\varpi_\mathrm{a}$} as a function of magnitude and shows that {the majority of} stars is at least 200 pc distant. Most measurements are within the $\pm 3\, \sigma_\varpi $ uncertainty  limits expected for stars located at infinite distance.

\begin{figure}
\includegraphics[width=\linewidth]{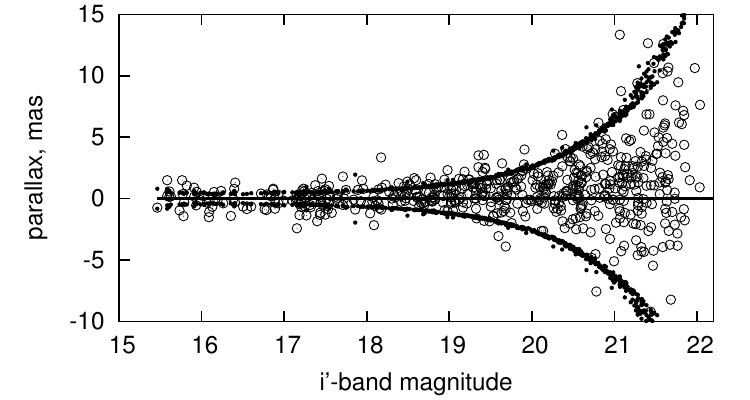}
\caption {The measured absolute parallaxes of field stars (open circles) which lie mostly within statistical  $+3\sigma$ upper and $-3\sigma$ lower limits indicated by solid circles for stars at infinite distance. Absolute parallaxes were obtained from the relative parallaxes given in the catalogue and the 0.50 mas correction.}
\label{parall}
\end{figure}

{However, about 30 stars have negative parallaxes beyond the $-3\,\sigma_{\varpi}$ level and as shown in Fig. \ref{xx} those are concentrated in a few isolated areas of the FoV periphery. Whereas underestimated parallax uncertainties are unlikely to lead to this space-correlated pattern, a slow change of the zero point of parallaxes across the FoV could be the cause. As commented in Sect. \ref{astr}, the parallaxes are determined on the basis of polynomial functions in the spatial coordinates  $x$,$y$. Therefore, if the spatial distribution of parallaxes is strongly inhomogeneous, the systematic difference between $\varpi_\mathrm{a}$ and the real parallaxes is not flat and can slowly change with $x$,$y$, creating local zones with deviating parallaxes. In Figure \ref{xx} we can see such areas where parallaxes of bright stars are excessively negative. The size of these areas is small and no outlying parallaxes are found in other parts of the FoV, in particular in the field centre. Therefore our estimate of the parallax of \dw\ is not expected to be biased. In addition, we verified that limiting the reference star sample for the parallax correction (Sect. \ref{sec:astroparam}) to stars within $<1\arcmin$, i.e.\ excluding the parallax outliers, yields an essentially identical result $\Delta \varpi = -0.44 \pm 0.10$.}

\begin{figure}
\includegraphics[width= \linewidth]{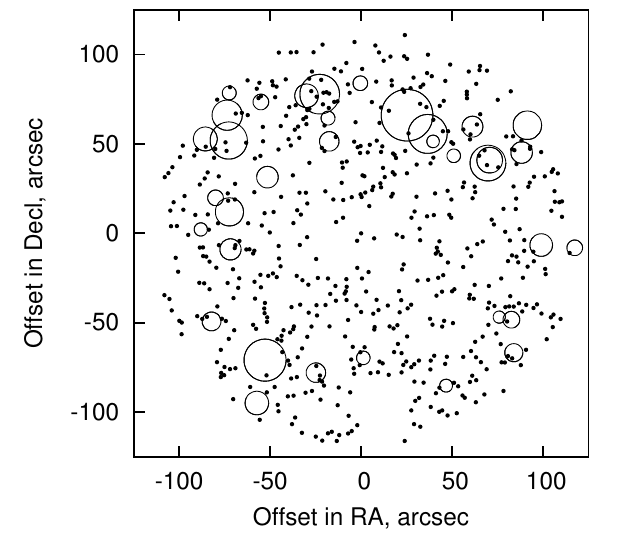} 
\caption{{On-sky position of all  catalogue stars (black dots) and stars which  measured absolute parallax is below $-3\,\sigma_{\varpi}$ (open circles). The circle size is proportional to the ratio $-\varpi_\mathrm{a} / \sigma_{\varpi}$.}}
\label{xx}
\end{figure}

The relative parallaxes and proper motions {of reference stars} are available at the CDS as a catalogue which contains RA and Dec in the ICRF. Transformation to the ICRF system was performed using the USNO-B catalogue \citep{USNO}.  There are 45 stars in the \dw\ field for which we obtained precise astrometry with OSIRIS and that were included in USNO-B. We mapped their positions using quadratic and cubic polynomials with a resulting r.m.s. difference of 0\farcs17, which allowed us to find  the absolute positions with an uncertainty of $\sim$ $0\farcs04 - 0\farcs2 $, propagated from the USNO-B data. This exceeds the precision of the original relative positions by a factor $\sim$1000. We plan to update the absolute positions using the \emph{Gaia} astrometric catalogues, when they become available. The $i'$-band magnitudes will have to be updated as well, because none of the reference stars has published $i'$-band photometry, necessary for conversion from the instrumental to standard magnitudes. The only reference is the Cousins $I_\mathrm{C}=17.0\pm0.2$ estimate given by \citet{Koen:2013uq} for \dw, which we used for zero-point determination. We neglected the difference between the Cousins $I_\mathrm{C}$ and Sloan $i'$ systems, which is small in comparison to the random error of 0.2 mag of the \citet{Koen:2013uq} measurement. Consequently, although the zero-point is from a $I_\mathrm{C}$ magnitude, our photometry relates to Sloan $i'$.

The catalogue also contains the quality flag $\chi^2$ for each star, which is the average ratio of $x_{\rm ep}^2$ and $y_{\rm ep}^2$ to the model variance. For a random sample following a normal distribution, we expect to register $\chi^2 \approx 1$ on average, but the measured data  in Fig. \ref{chi2} exhibits a large scatter and an increase at the bright end. The observed large deviations $\chi^2 \geq 10$ were found to correspond to blended images. Excluding those cases, we found a linear dependence of $\chi^2$ on $i'$, which approaches unity for faint stars and a value of 2.6 at the magnitude of \dw. The value $\chi^2=4.5$ obtained for \dw\ lies in the tail of the $\chi^2$ distribution for field stars. A smaller excess in $\chi^2$, usually within $1.1-1.5$ {and not related to underestimated uncertainties}, was already noticed for FORS2 observations \citep{Lazorenko:2014aa}. The OSIRIS data do not allow us to conclude on the nature of the systematic excess in $\chi^2$ and we attribute it to a systematic error that remains to be characterised.

Table \ref{tab:cat} shows the content of the catalogue, which contains the sequential star number Nr in the field, the $i'$-band magnitude and its internal precision $\sigma_{i'}$,  RA and Dec for the equinox and epoch J2000.0 with their uncertainties, the relative proper motion $\mu_{\alpha^\star}$,$\cos \delta$ and $\mu_{\delta}$ per Julian year and uncertainties, the relative parallax $\varpi$ with uncertainty $\sigma_\varpi$, and the DCR parameter $\rho$. The mean epoch of observations is given by $\bar T$  in Julian years {since J2000}, and the $\chi^2$-value for the epoch residuals flags the quality of the fit. 

\begin{table*}
\caption{Excerpt of the astrometric catalogue. The full table is available online and at the CDS.}
\centering
\tabcolsep=0.11cm
\begin{tabular}{ccccccccccccccccccc}
\hline
Nr  &   $i'$   &$\sigma_{i'}$& RA &$\sigma_{\alpha}$  & Dec                 &$\sigma_{\delta}$&$\mu_{\alpha^\star}$&$\sigma_{\mu_{\alpha^\star}}$ &$\mu_{\delta}$&$\sigma_{\mu_{\delta}}$&$\rho$&$\sigma_\rho$&$\varpi$&$\sigma_\varpi$&$\bar T$&$\chi^2$\\
           & (mag)  &    (mag)      &  (deg)  & (\arcsec)                      &(deg)             &(\arcsec)                   &      (mas                 & (mas                   & (mas & (mas & (mas) & (mas)&(mas)&(mas)&(yr)  &        \\
         &  &       &   &            &                    &                   &      /yr)                 &  /yr)          & /yr) & /yr) & &  &   \\
\hline
41 & 20.409 & 0.006 & 275.3611854 & 0.639 & 14.2014463 & 0.537 & -0.9 & 1.16 & 0.39 & 0.91 & 3.28 & 0.94 & 3.53 & 1.21 & 14.0043 & 0.63 \\
51 & 20.061 & 0.005 & 275.3601554 & 0.604 & 14.2020711 & 0.506 & -0.26 & 0.85 & 3.12 & 0.67 & -4.46 & 0.71 & 0.63 & 0.92 & 14.0043 & 0.81 \\
54 & 20.597 & 0.006 & 275.3638648 & 0.579 & 14.2023064 & 0.484 & -3.97 & 1.36 & -4.53 & 1.06 & -4.34 & 1.17 & -1.07 & 1.44 & 14.0043 & 2.14 \\
58 & 19.437 & 0.004 & 275.3652450 & 0.561 & 14.2025570 & 0.468 & 5.1 & 0.51 & -3.38 & 0.39 & 6.82 & 0.42 & 1.92 & 0.54 & 14.0043 & 3.94 \\
60 & 18.418 & 0.002 & 275.3620758 & 0.567 & 14.2025922 & 0.473 & 1.63 & 0.25 & 1.25 & 0.19 & -0.49 & 0.2 & -0.74 & 0.27 & 14.0043 & 2.96 \\
\hline
\end{tabular}
\label{tab:cat}
\end{table*}

\begin{figure}
\includegraphics[width=\linewidth]{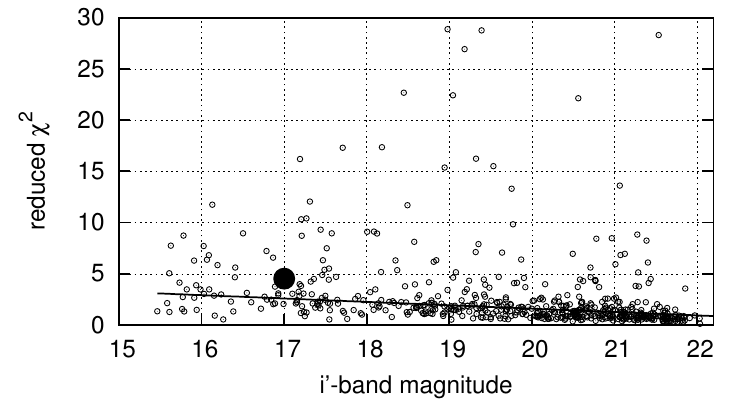}
\caption {Distribution of the reduced $\chi^2$ values as a function of magnitude for field stars (open circles) and a linear approximation (solid line). The solid circle marks \dw\ with $\chi^2=4.5$.}
\label{chi2}
\end{figure}

\subsection{OSIRIS pixel scale}{\label{scale}}
The conversion from CCD pixel space to on-sky angular space was performed based on the USNO-B positions. This allowed us to derive a pixel scale of $129.11 \pm 0.18$~mas/px, which differs from the value of 127~mas/px given in the OSIRIS documentation (\url{http://www.gtc.iac.es/instruments/osiris/osiris.php}). This difference could be caused by the thermal change of the telescope focal length and we emphasise that our updated scale is valid at the target position only. The difference between the pixel scales we determined along $x$ and $y$ axes is $0.16 \pm 0.25$~mas/px, thus not significant.

\section{Conclusions}
We obtained multi-epoch imaging observations of the L4.5 dwarf \dw\ with OSIRIS/GTC in $i`$-band spanning 17 months and applied methods developed for FORS2/VLT to measure its astrometric motion relative to field stars. We performed a first analysis of the astrometric performance of OSIRIS in comparison with the well-studied properties of FORS2. The segmented structure of GTC's main mirror produces a slightly more complicated PSF shape, which however does not significantly affect the astrometry. 

With OSIRIS we achieved a single-frame astrometric precision of $1.0$ mas for a well-exposed star, an exposure time of 45 s, the reference star density of the \dw\ field, and the average FWHM of $0\farcs80$. The expected precision for one epoch consisting of 28 single frames is then 0.23 mas. However, the measured r.m.s. dispersion of epoch residuals is $\sigma_{\rm ep}=0.35-0.40$ mas, thus significantly larger for reasons not yet fully understood. With additional OSIRIS data taken with a similar observation strategy, we hope to develop optimised reduction methods to reach the 0.23 mas accuracy limit set by the image motion, seeing, and the reference frame noise. 

In comparison to typical FORS2 observation at VLT presented in \cite{Lazorenko:2014aa} and \cite{Sahlmann:2014aa}, we have measured a 23\% larger FWHM and a factor of 2.2 larger image motion for the \dw\ observations with OSIRIS. These two factors explain why the photocentre precision is 0.23 mas for OSIRIS compared to 0.12 mas for FORS2. The larger image motion of OSIRIS can be explained by larger atmospheric turbulence at high altitudes, by instabilities of the GTC and its optics, or by a combination thereof. GTC observations in better atmospheric conditions than met here are expected to yield better astrometric precision. 

Using the eleven observation epochs, we determined a trigonometric distance of $9.38 \pm 0.03$ pc to the L4.5 dwarf \dw, which represents the first astrophysical application of precision astrometry with OSIRIS/GTC. This is also the first parallax determination for \dw, which establishes it as a member of the 10 pc sample. We measured the proper motion of \dw\ with high-precision and the resulting galactic kinematics are consistent with the suspected youth of this source. The data exclude the presence of binary or planetary companions with masses as low as 5 $M_J$ and periods of 50--1000 days ($\approx$0.1--0.7 au), which illustrates the potential of OSIRIS astrometry for exoplanet and binary search and orbit characterisation.

In summary, we demonstrated that the OSIRIS camera of the GTC is capable of measuring differential positions with an accuracy of 0.35--0.40 mas over a timespan of 1.4 years. We thus established OSIRIS as an instrument suitable for high-precision astrometry for faint optical sources located in the northern hemisphere. This study is relevant for the development of astrometric programs with extremely large optical telescopes like the E-ELT and TMT, which also employ segmented primary mirrors. The precision level of 0.1--0.3 mas explored here is however one order of magnitude larger than what can be expected for optical telescopes with apertures of 30--40 m. 

\section*{Acknowledgments}
J.S. is supported by an ESA Research Fellowship in Space Science. This research made use of the databases at the Centre de Donn\'ees astronomiques de Strasbourg (\url{http://cds.u-strasbg.fr}); of NASA's Astrophysics Data System Service (\url{http://adsabs.harvard.edu/abstract\_service.html}); of the paper repositories at arXiv; of the M, L, T, and Y dwarf compendium housed at \url{DwarfArchives.org}; and of Astropy, a community-developed core Python package for Astronomy \citep{Astropy-Collaboration:2013aa}.

\bibliographystyle{mn2e_spec}
\bibliography{pa} 
\bsp
\label{lastpage}
\end{document}